\documentclass[twocolumn]{aastex62}

\usepackage{ulem}
\hypersetup{linkcolor=red,citecolor=blue,filecolor=cyan,urlcolor=magenta}
\usepackage{amsmath}

\newcommand{\MO}{$M_\odot$}
\newcommand{\ZO}{$Z_\odot$}
\newcommand{\Chandra}{\textit{Chandra}}
\newcommand{\XMM}{\textit{XMM-Newton}}
\newcommand{\HST}{\textit{HST}}

\newcommand{\rxj}{RX\,J1347.5--1145}

\received{}
\revised{}
\accepted{}
\submitjournal{ApJ}

\shorttitle{Cool core disturbed: sub-sonic sloshing gas and stripped shock-heated gas}
\shortauthors{Ueda et al.}

\begin{document}

\title{Cool core disturbed: Observational evidence for coexistence of sub-sonic sloshing gas and stripped shock-heated gas around the core of \rxj}


\correspondingauthor{Shutaro Ueda}
\email{sueda@asiaa.sinica.edu.tw}
\author[0000-0001-6252-7922]{Shutaro Ueda}
\affiliation{Academia Sinica Institute of Astronomy and Astrophysics (ASIAA), No. 1, Section 4, Roosevelt Road, Taipei 10617, Taiwan}
\affiliation{Institute of Space and Astronautical Science (ISAS), Japan Aerospace Exploration Agency (JAXA), 3-1-1 Yoshinodai, Chuo, Sagamihara, Kanagawa 252-5210, Japan}

\author{Tetsu Kitayama}
\affiliation{Department of Physics, Toho University, 2-2-1 Miyama, Funabashi, Chiba 274-8510, Japan}

\author{Masamune Oguri}
\affiliation{Research Center for the Early Universe, School of Science, 
The University of Tokyo, 7-3-1 Hongo, Bunkyo, Tokyo 113-0033, Japan}
\affiliation{Department of Physics, The University of Tokyo, 7-3-1 Hongo, Bunkyo, Tokyo 113-0033, Japan}
\affiliation{Kavli Institute for the Physics and Mathematics of the Universe (Kavli IPMU), The University of Tokyo, 5-1-5 Kashiwanoha, Kashiwa, Chiba 277-8583, Japan}

\author{Eiichiro Komatsu}
\affiliation{Kavli Institute for the Physics and Mathematics of the Universe (Kavli IPMU), The University of Tokyo, 5-1-5 Kashiwanoha, Kashiwa, Chiba 277-8583, Japan}
\affiliation{Max-Planck-Institut f\"{u}r Astrophysik, Karl-Schwarzschild Str. 1, D-85741 Garching, Germany}

\author{Takuya Akahori}
\affiliation{Mizusawa VLBI observatory, National Astronomical Observatory of Japan, 2-21-1 Osawa, Mitaka, Tokyo 181-8588, Japan}

\author{Daisuke Iono}
\affiliation{National Astronomical Observatory of Japan, 2-21-1 Osawa, Mitaka, Tokyo 181-8588, Japan}
\affiliation{The Graduate University for Advanced Studies (SOKENDAI), 
2-21-1 Osawa, Mitaka, Tokyo 181-8588, Japan}

\author{Takumi Izumi}
\affiliation{National Astronomical Observatory of Japan, 2-21-1 Osawa, Mitaka, Tokyo 181-8588, Japan}

\author{Ryohei Kawabe}
\affiliation{National Astronomical Observatory of Japan, 2-21-1 Osawa, Mitaka, Tokyo 181-8588, Japan}
\affiliation{The Graduate University for Advanced Studies (SOKENDAI), 2-21-1 Osawa, Mitaka, Tokyo 181-8588, Japan}
\affiliation{Department of Astronomy, 
The University of Tokyo, 7-3-1 Hongo, Bunkyo, Tokyo 113-0033, Japan}

\author{Kotaro Kohno}
\affiliation{Research Center for the Early Universe, School of Science, The University of Tokyo, 7-3-1 Hongo, Bunkyo, Tokyo 113-0033, Japan}
\affiliation{Institute of Astronomy, The University of Tokyo, 2-21-1 Osawa, Mitaka, Tokyo 181-0015, Japan}

\author{Hiroshi Matsuo}
\affiliation{National Astronomical Observatory of Japan, 2-21-1 Osawa, Mitaka, Tokyo 181-8588, Japan}
\affiliation{The Graduate University for Advanced Studies (SOKENDAI), 2-21-1 Osawa, Mitaka, Tokyo 181-8588, Japan}

\author{Naomi Ota}
\affiliation{Department of Physics, Nara Women's University, Kitauoyanishi-machi, Nara, Nara 630-8506, Japan}

\author{Yasushi Suto}
\affiliation{Research Center for the Early Universe, School of Science, The University of Tokyo, 7-3-1 Hongo, Bunkyo, Tokyo 113-0033, Japan}
\affiliation{Department of Physics, The University of Tokyo, 7-3-1 Hongo, Bunkyo, Tokyo 113-0033, Japan}

\author{Shigehisa Takakuwa}
\affiliation{Department of Physics and Astronomy, Graduate School of Science and Engineering,
Kagoshima University, 1-21-35 Korimoto, Kagoshima, Kagoshima 890-0065, Japan}
\affiliation{Academia Sinica Institute of Astronomy and Astrophysics (ASIAA), No. 1, Section 4, Roosevelt Road, Taipei 10617, Taiwan}

\author{Motokazu Takizawa}
\affiliation{Department of Physics, Yamagata University, 1-4-12 Kojirakawa-machi, Yamagata, Yamagata 990-8560, Japan}

\author{Takahiro Tsutsumi}
\affiliation{National Radio Astronomy Observatory, P.O. Box O, Socorro, NM, 87801, USA}

\author{Kohji Yoshikawa}
\affiliation{Center for Computational Sciences, University of Tsukuba, 1-1-1 Tennodai, Tsukuba, Ibaraki 305-8577, Japan}

%



\begin{abstract}
\rxj ~($z=0.451$) is one of the most luminous X-ray galaxy clusters, which hosts a prominent cool core and exhibits a signature of a major merger. We present the first {\it direct} observational evidence for sub-sonic nature of sloshing motion of the cool core. We find that a residual X-ray image from the {\it Chandra X-ray Observatory} after removing the global emission shows a clear dipolar pattern characteristic of gas sloshing, whereas we find no significant residual in the Sunyaev-Zel'dovich effect (SZE) image from the Atacama Large Millimeter/submillimeter Array (ALMA). We estimate the equation of state of perturbations in the gas from the X-ray and SZE residual images. The inferred velocity is $420^{+310}_{-420}$\,km\,s$^{-1}$, which is much lower than the adiabatic sound speed of the intracluster medium in the core. We thus conclude that the perturbation is nearly isobaric, and gas sloshing motion is consistent with being in pressure equilibrium. Next, we report evidence for gas stripping of an infalling subcluster, which likely shock-heats gas to high temperature well in excess of 20\,keV. Using mass distribution inferred from strong lensing images of the {\it Hubble Space Telescope} (\HST), we find that the mass peak is located away from the peak position of stripped gas with statistical significance of $> 5\sigma$. Unlike for the gas sloshing, the velocity inferred from the equation of state of the excess hot gas is comparable to the adiabatic sound speed expected for the 20\,keV intracluster medium. All of the results support that the southeast substructure is created by a merger. On the other hand, the positional offset between the mass and the gas limits the self-interaction cross section of dark matter to be less than $3.7~h^{-1}~{\rm cm^2~g^{-1}}$ (95\%~CL). 
\end{abstract}


\section{Introduction}

Galaxy clusters are the largest gravitationally-bound and virialized objects in the universe. They are located at the knots of filaments in the large-scale structure and provide us with unique cosmological information. Galaxy clusters are also dynamically young and are continuously growing through mergers between smaller clusters. Such merger activities induce various high-energy phenomena in the hot and optically thin plasma, i.e., the intracluster medium (ICM), in the gravitational potential well of clusters. 

A large amount of observational and numerical studies have suggested that mergers lead to shock-heating of the gas \citep[e.g.,][]{Ricker01, Markevitch02, Takizawa05, Takizawa06, Bourdin13}, production of cold fronts \citep[e.g.,][]{Markevitch01, Ascasibar06, ZuHone10, Roediger11, Blanton11, Ueda17, Hitomi18d}, ram-pressure stripping of the gas from infalling galaxies \citep[e.g.,][]{David04, Roediger07, Sasaki16}, non-equilibrium ionization of the ICM \citep[e.g.,][]{Akahori08, Akahori10, Akahori12, Inoue16}, and (re-) acceleration of relativistic particles \citep[e.g.,][]{Feretti12, Akamatsu13b, Brunetti14, van_Weeren17}. The entire picture of merging processes is however still far from clear; for example, heating mechanisms and dynamics of the ICM are under debate. Observational studies in multi-wavelength are therefore crucial for understanding the physics of galaxy cluster mergers.

\rxj ~is one of the most luminous X-ray galaxy clusters and is located at the redshift of $z = 0.451$. It was thought to be a relaxed cluster when it was discovered in the {\it ROSAT} all sky survey \citep{Schindler97}. \cite{Komatsu99} made the first measurements of the Sunyaev-Zel'dovich effect  \citep[SZE:][]{Sunyaev72} toward this cluster with the James Clerk Maxwell Telescope (JCMT) at 350\,GHz as well as with the 45\,m Nobeyama Radio Telescope at 21 and 43\,GHz. A higher angular resolution observation of the SZE was performed by \cite{Komatsu01} using the Nobeyama Bolometer Array (NOBA) and they found a prominent substructure which has no counterpart in the soft X-ray image by {\it ROSAT}. The presence of the substructure has been confirmed by \Chandra ~and \XMM ~\citep[e.g.,][]{Allen02, Gitti04} as well as by more recent SZE measurements \citep[][]{Mason10, Korngut11, Plagge13, Adam14, Kitayama16}. \cite{Allen02} measured the mean temperature of the ICM to be over 10\,keV, which is relatively high compared to other typical clusters. \cite{Kitayama04} and \cite{Ota08} found a very hot ($> 20$\,keV) component of the ICM in this cluster. In addition, the radial profile and spatial distribution of the ICM temperature indicate that the temperature drops to $\sim 6$\,keV toward the cluster center so that the cool core is formed \citep[e.g.,][]{Allen02, Ota08, Kreisch16}. A disturbed morphology is furthermore supported by the radio synchrotron observations \citep[e.g.,][]{Ferrari11} and the gravitational lensing maps \citep[e.g.,][]{Kohlinger14}. The total mass of \rxj ~within $r_{200}$ is estimated to be $\sim 1.5 \times 10^{15}\,h^{-1}$\,\MO ~by using the weak-lensing analysis, where $r_{200}$, the radius within which the mean mass density is 200 times the critical density of the universe, is $1.85\,h^{-1}$\,Mpc \citep{Lu10} for this galaxy cluster\footnote{They adopted the Hubble constant of 70\,km\,s$^{-1}$\,Mpc$^{-1}$.}.

It is considered that the hot component is most likely associated with a past major merger event \citep[e.g.,][]{Johnson12, Kreisch16}, although its specific nature, such as geometry and dynamics of the collision, is still unclear. Recently, \cite{Kitayama16} presented the SZE image observed by Atacama Large Millimeter/submillimeter Array (ALMA) with angular resolution of 5\arcsec. Such high angular resolution enables us to remove the emission of the central active galactic nucleus (AGN) and to reconstruct an accurate SZE map. \cite{Kitayama16} found that the shape of the SZE is elongated toward the southeast and the peak position of the SZE is located at 11\arcsec ~southeast from the central AGN. The ALMA high-resolution SZE image motivates us to directly compare it with high-quality data in other wavelengths by \Chandra ~and {\it Hubble Space Telescope} (\HST). In this paper, we investigate the merger phenomena and merger history in \rxj ~by combining the data of \Chandra, ALMA, and \HST. 

We adopt $\Omega_{\rm m}=0.3$ and $\Omega_{\rm \Lambda}=0.7$. We use the dimensionless Hubble constant ($h \equiv H_{0}/100\,$km\,s$^{-1}$\,Mpc$^{-1}$); given controversial results on the value of $h$ \citep[e.g.,][]{Planck16, Riess16}, we do not fix it unless stated otherwise. In this cosmology, an angular size of $1''$ corresponds to a physical scale of 4.04\,$h^{-1}$\,kpc at the redshift $z=0.451$. Unless stated otherwise, quoted errors correspond to 1$\sigma$.

\section{Observations and Data Reductions}
\label{sec:obs}

We used the X-ray data of \rxj ~taken with the Advanced CCD Imaging Spectrometer \citep[ACIS;][]{Garmire03} on board \Chandra. Two out of six datasets we used were taken by ACIS-S (ObsID: 506 and 507) and the others were by ACIS-I (ObsID: 3592, 13516, 13999, and 14407). All the datasets are the same as used in our previous study \citep{Kitayama16}. The data reduction is also the same as in \cite{Kitayama16}, except that we used the updated versions, 4.9 and 4.7.5.1, of \Chandra ~Interactive Analysis of Observations \citep[CIAO;][]{Fruscione06} and the calibration database (CALDB), respectively. The background data were extracted from the region between $2.5'$ and $3.5'$ from the peak position of the cluster. Unless stated otherwise, we used \texttt{XSPEC} version 12.9.1$l$ \citep{Arnaud96} and the atomic database for plasma emission modeling (AtomDB) version 3.0.9 in the X-ray spectral analysis, assuming that the ICM is in a collisional ionization equilibrium. The Galactic absorption was fixed at $N_{\rm H} = 4.6 \times 10^{20}$\,cm$^{-2}$ \citep{Kalberla05}. We adopted the abundance table of \cite{Anders89} in the X-ray spectral analysis. 

The ALMA SZE data used in this paper are the same as those in \cite{Kitayama16}, except that an updated version, 5.1.1, of the Common Astronomy Software Applications \citep[{\tt CASA}:][]{McMullin07} was adopted in the imaging analysis.

We also used the data of \HST ~observations of \rxj ~obtained with the Advanced Camera for Surveys (ACS) to study the central mass distribution from strong lensing (SL). These data were provided by the Space Telescope Science Institute (STScI) and no additional data reduction was applied in this study. The data consisted of F475W, F814W, and F850LP images from the ACS, which were used in \cite{Halkola08}, \cite{Kohlinger14}, and \cite{Zitrin15}.

\section{X-ray and SZE Analyses}
\label{sec:x}

\subsection{X-ray and SZE imaging analysis}
\label{sec:ximage}

\begin{figure*}
 \begin{center}
  \includegraphics[width=8.5cm]{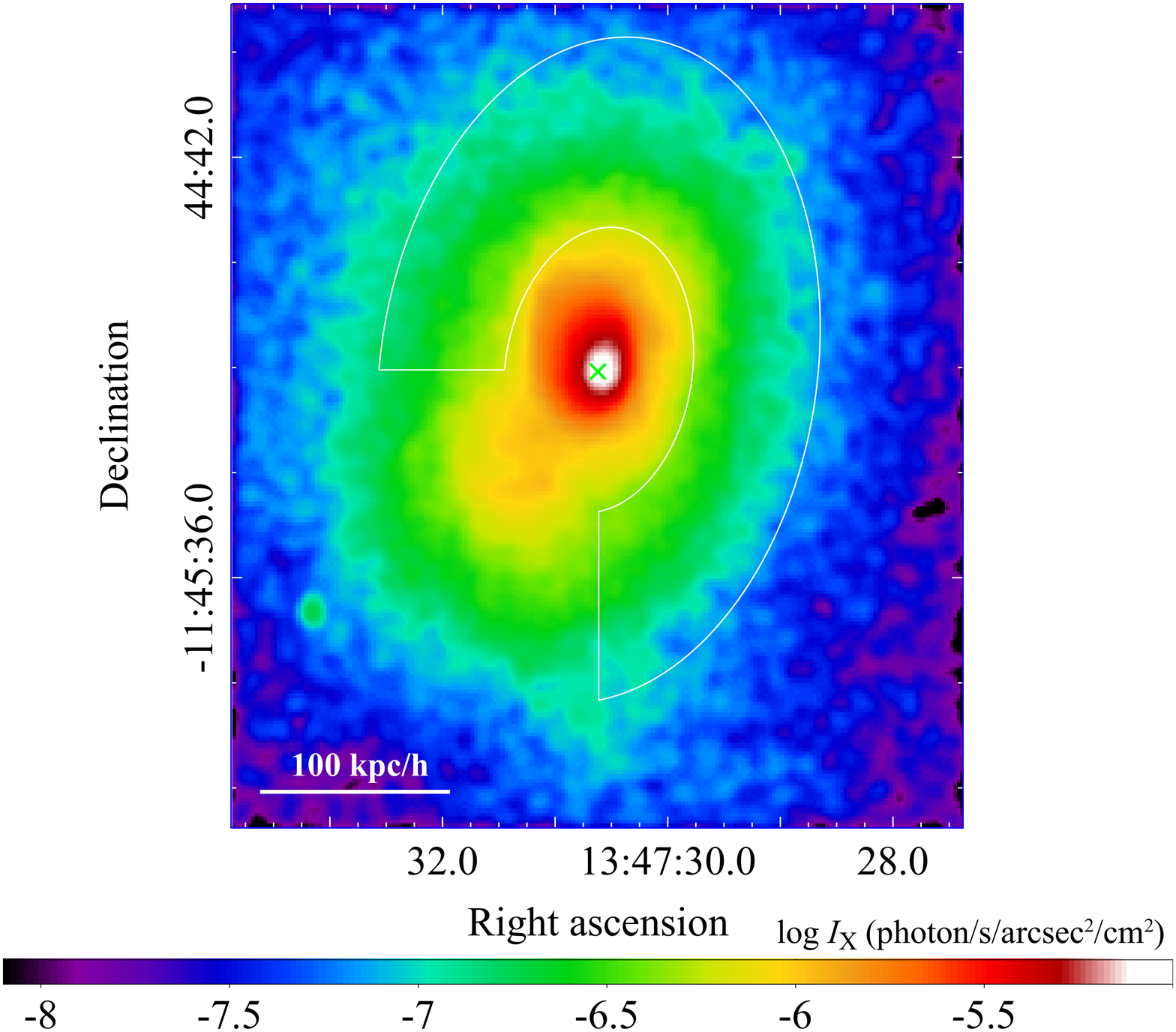}
  \includegraphics[width=8.5cm]{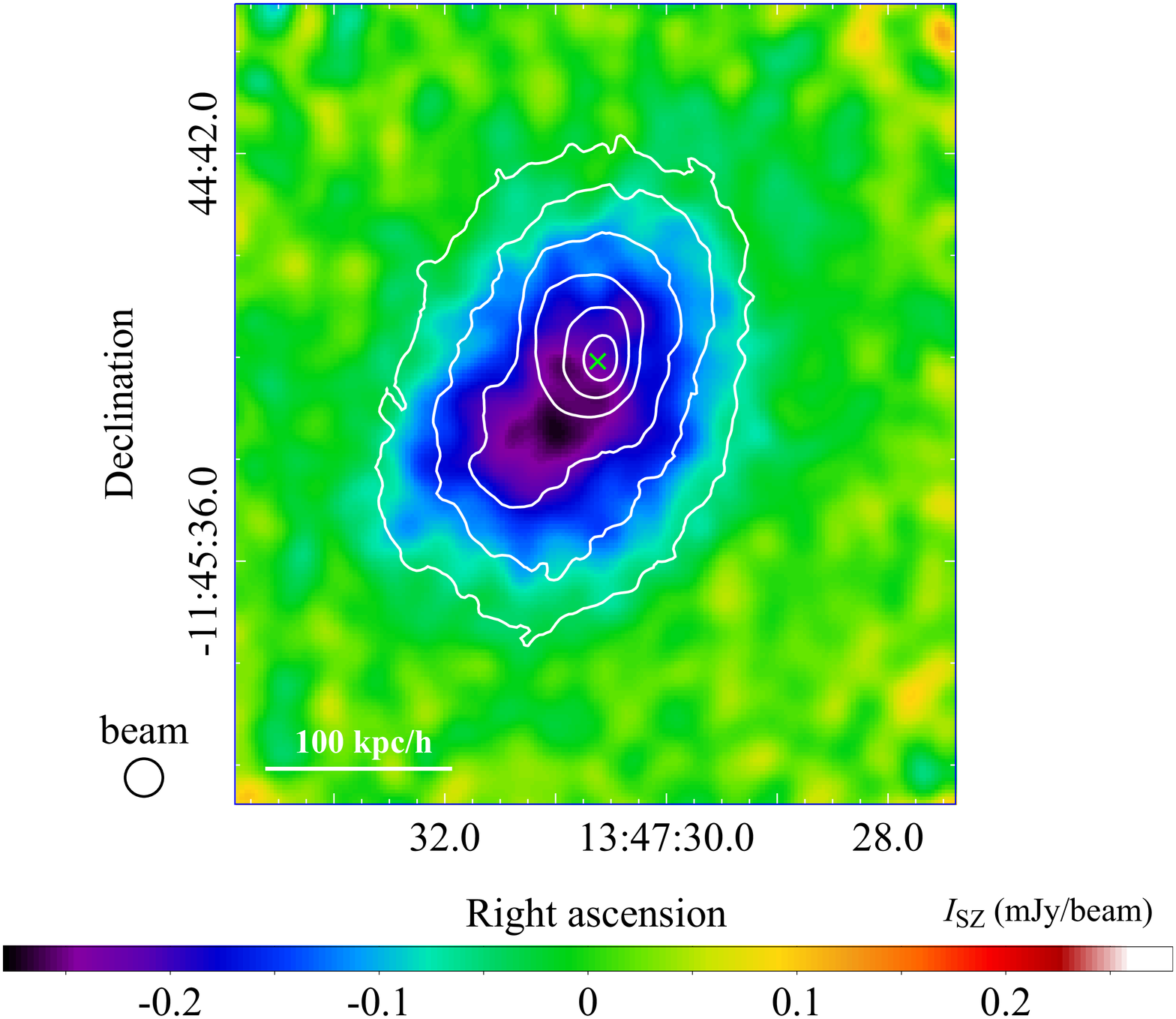} 
 \end{center}
\caption{Surface brightnesses of the X-ray ({\it left}) and the SZE ({\it right}) of \rxj. The green cross corresponds to the position of the central AGN identified by ALMA at 92\,GHz. {\it Left:} The X-ray surface brightness in the $0.4 - 7.0$\,keV band is shown as the logarithm value in units of photon\,sec$^{-1}$\,arcsec$^{-2}$\,cm$^{-2}$. This image is smoothed by the Gaussian kernel with $2.3''$ FWHM. Overlaid is a white elliptical annulus within which variation of this surface brightness is minimized at the mean radius between $15''$ and $35''$, excluding the southeast quadrant.
{\it Right:} The SZE surface brightness overlaid with the contours of the X-ray surface brightness, corresponding to 64, 32, 18, 8, 4, and 2\,$\%$ of the peak value, after being smoothed by the Gaussian kernel with $2.3''$ FWHM. Colors indicate the ALMA SZE map in mJy\,beam$^{-1}$ smoothed to a beam size of $5''$ FWHM. The beam size is presented as a black circle out of the map. 
}
\label{fig:sb}
\end{figure*}

\begin{figure*}
 \begin{center}
  \includegraphics[width=8.5cm]{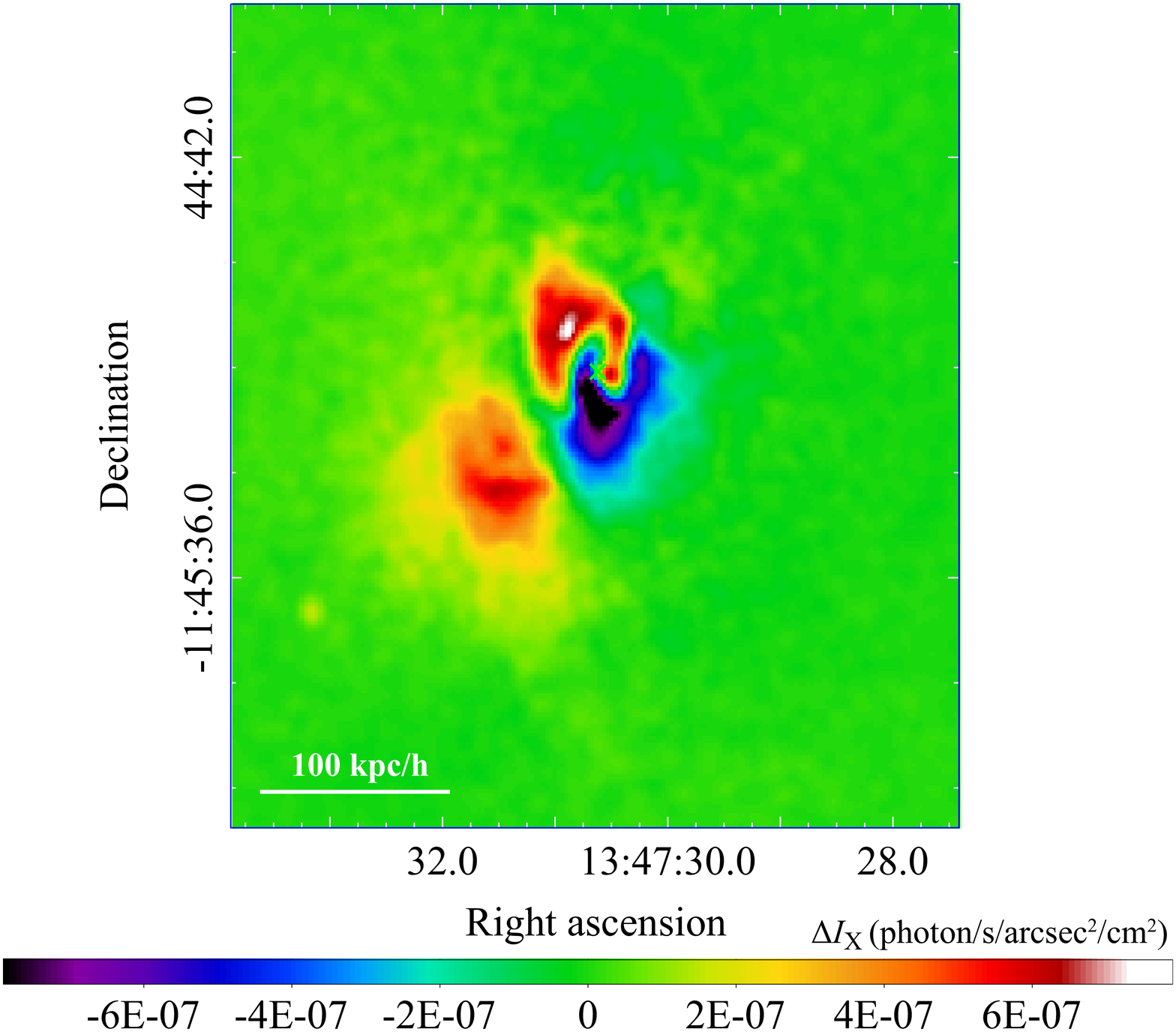}
  \includegraphics[width=8.5cm]{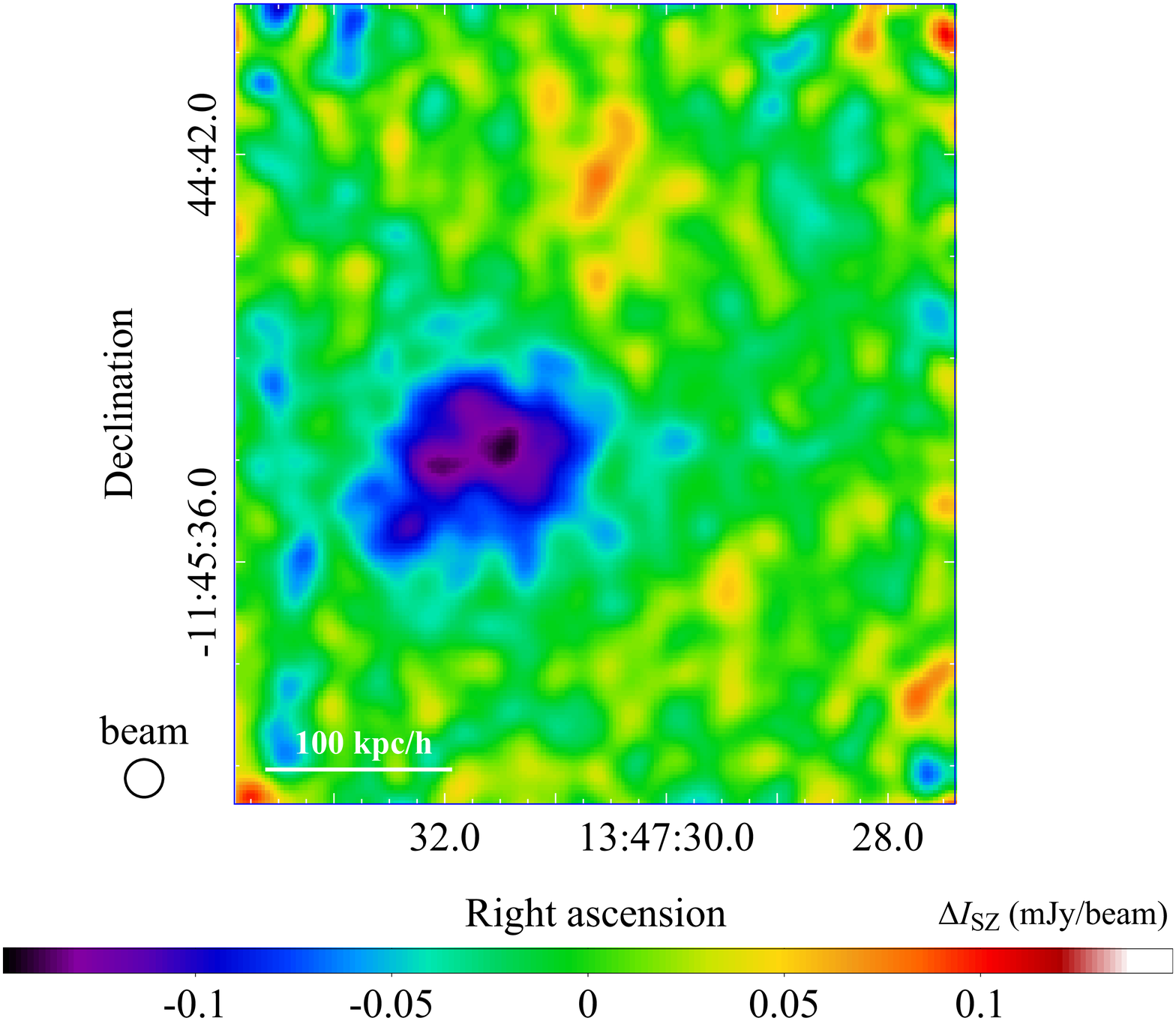} 
 \end{center}
\caption{Residual images of the X-ray and SZE surface brightness. 
{\it Left:} The residual image of \Chandra ~X-ray surface brightness after subtracting the mean profile, excluding the southeast quadrant. The original surface brightness is the same as the left panel of Figure~\ref{fig:sb}. For reference, the position of the central AGN is marked by a cross.
{\it Right:} Same as the left panel but for the ALMA~92\,GHz brightness in mJy\,beam$^{-1}$ with the effective beam size of $5''$ FWHM. The original surface brightness is shown in the right panel of Figure~\ref{fig:sb}.
}
\label{fig:sbellip_xsz}
\end{figure*}

Figure~\ref{fig:sb} shows the X-ray surface brightness ({\it left}) and the SZE image ({\it right}) of \rxj, respectively, both of which clearly exhibits a substructure in the southeast direction. First, we computed the mean X-ray surface brightness over an ellipse, excluding the southeast quadrant (i.e., the substructure), following \cite{Ueda17}. To accurately model the mean surface brightness profile, we searched for the ellipse that minimizes the variance of the X-ray surface brightness relative to its mean in a $15'' < \bar{r} < 35''$ annulus, where $\bar{r}$ is the geometrical mean of the semi-major and semi-minor axis lengths around the X-ray peak. The range of $\bar{r}$ was chosen to match the position of the excess emission in the excluded southeast quadrant and to eliminate the impact of the bright core ($\bar{r}<10''$) of this cluster on the overall shape.  We find that the axis ratio of the ellipse is 0.66 and its position angle is $-8.7^{\circ}$ \footnote{Throughout this paper, the position angle is measured for the major axis of an ellipse from north (0\,deg) through east (90\,deg).}. The resultant annulus is shown in the left panel of Figure~\ref{fig:sb}. We then calculated the mean X-ray brightness over the ellipse at each $\bar{r}$, excluding the southeast quadrant, and subtracted it from the entire brightness. The left panel of Figure~\ref{fig:sbellip_xsz} shows an X-ray residual image of \rxj.

We also applied the same procedure for the SZE image. The obtained SZE residual image is shown in the right panel of Figure~\ref{fig:sbellip_xsz}. Note that we used the unsmoothed SZE image with the synthesized beam size of $4.1'' \times 2.5''$ FWHM \citep{Kitayama16} in the SZE imaging analysis. The smoothed images are shown only for display purposes. 

Figure~\ref{fig:sbellip_xsz} clearly shows an excess signal in the southeast quadrant both in the X-ray and SZE images. Thanks to the high-resolution data, we find that the excess SZE signal has the average FWHM of $\sim 28''$ (113\,$h^{-1}$\,kpc), which is significantly more extended than that of $\sim 17''$ (69\,$h^{-1}$\,kpc) in the X-ray residual image. This indicates that a high temperature region has a larger extent than seen in X-rays. We will present more detailed analyses and results on this region using two independent methods in Section~\ref{sec:X-ray_spec} and Section~\ref{sec:se_image}.

A dipolar pattern around the central AGN is apparent only in the X-ray residual image (the left panel of Figure~\ref{fig:sbellip_xsz}). The dipolar pattern consists of a northern positive excess and a southern negative excess. The fraction of each component against the mean brightness is roughly 20\,\%. The dipolar pattern is clearly distinct from the excess in the southeast quadrant and extends over a spatial scale of $\sim 100$\,$h^{-1}$\,kpc. We will describe more detailed results in Section~\ref{sec:sloshing}.

We assessed the uncertainty in subtracting the mean profile of the X-ray and SZE images by the following two methods. One is the same method as described above, except that the subtraction of the mean profile is done after smoothing the X-ray image to the same angular resolution as the ALMA image. The other is that we assumed circular symmetry for the mean profile of X-rays and SZE instead of the ellipse. In this check, we do not use any smoothed images. We then find that 1) a difference of angular resolution between \Chandra ~and ALMA does not affect the residual images, and 2) the geometry of the mean profile does not affect the substructure in the southeast quadrant, while it slightly modifies the shape of the dipolar pattern in the core.

\subsection{X-ray spectral analyses of the southeast substructure}
\label{sec:X-ray_spec}

\begin{figure*}
 \begin{center}
  \includegraphics[width=8.5cm]{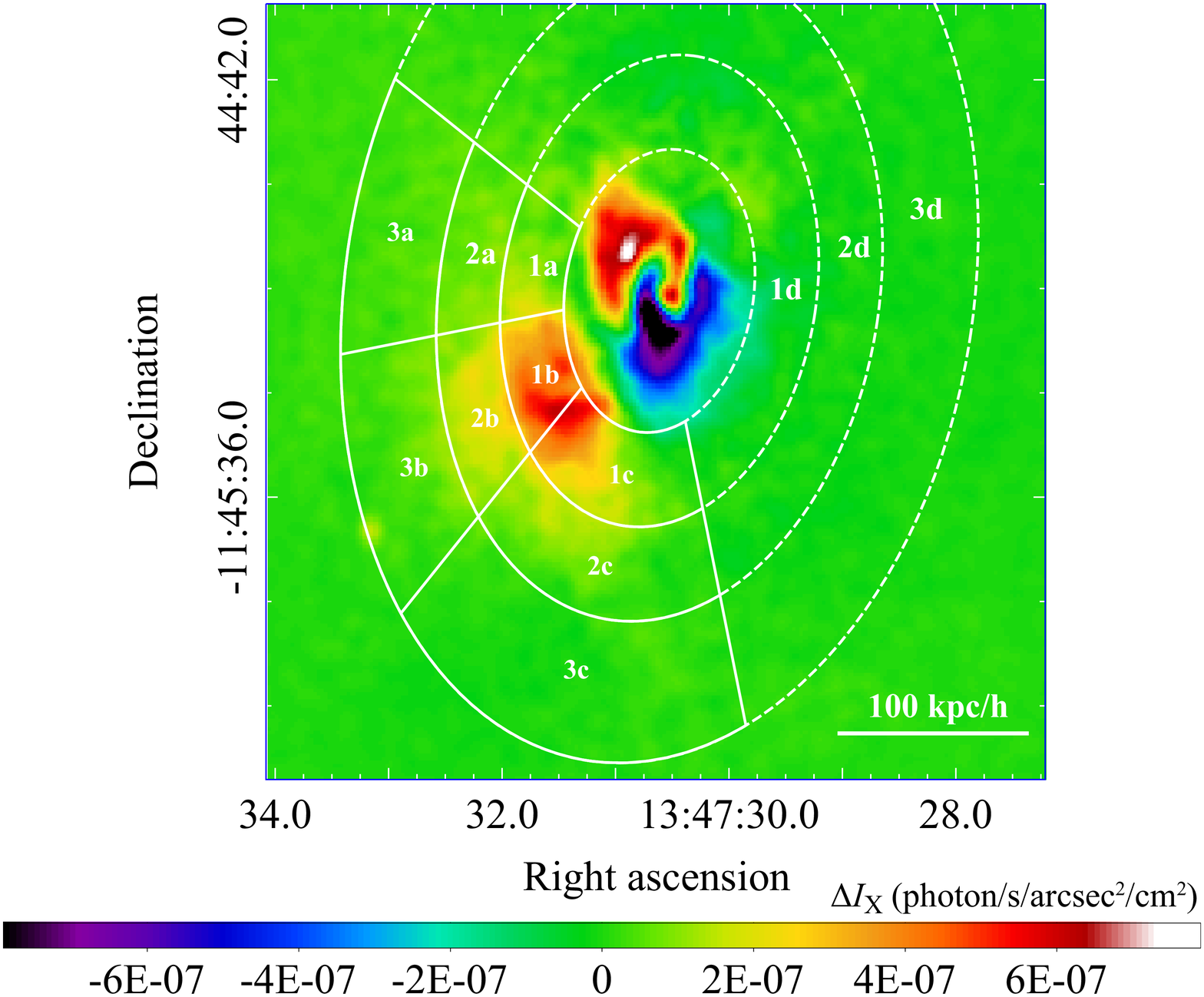} 
  \includegraphics[width=8.5cm]{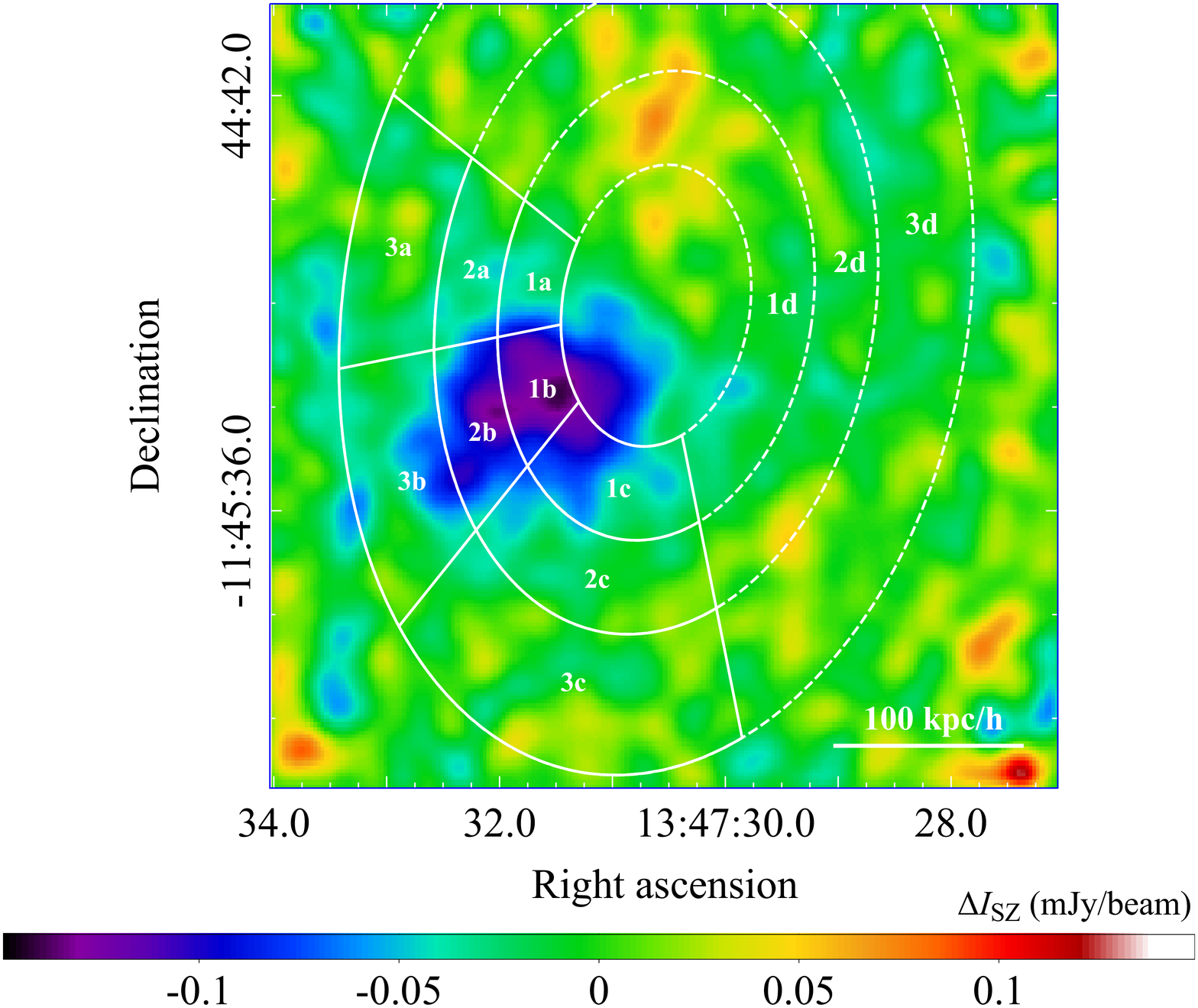} 
 \end{center}
\caption{Same as Figure \ref{fig:sbellip_xsz}, except that elliptical annuli used in the X-ray spectral analysis are overlaid.}
\label{fig:sbellip_9reg}
\end{figure*}

To understand the origin and thermodynamic properties of the ICM in the southeast substructure, we performed X-ray spectral analyses. We defined nine regions covering the excess and its surrounding along the ellipse as in Section~\ref{sec:ximage} (i.e., Region~1a, 1b, 1c, 2a, $\cdots$, 3b, and 3c) and three regions covering the remaining parts of the ellipse (i.e., Region 1d, 2d, and 3d) as illustrated in Figure~\ref{fig:sbellip_9reg}. First, we assumed, for simplicity, that the ICM in each region consists only of a single component over the entire line-of-sight. Its temperature ($kT_{\rm single}$) was measured using a model \texttt{phabs * apec} in \texttt{XSPEC}, where \texttt{phabs} represents the Galactic absorption \citep{Balucinska-Church92}. The photon counts in $0.4 - 7.0$\,keV and the best-fit parameters are summarized in Table~\ref{tab:se}.

Second, we adopted a two-component model in which an excess component is embedded in an ambient component in the nine regions.  The ambient component was assumed to have the same temperature as the gas temperature in an annulus at the same $\bar{r}$ (i.e., $kT_{\rm single}$ of Region~1d, 2d, and 3d). We then measured the temperature of the excess component ($kT_{\rm excess}$) by using the two \texttt{apec} model, fixing the temperature of the ambient component at the best-fit value of $kT_{\rm single}$ in Region~1d, 2d, and 3d, respectively. The spectral normalization of the ambient component was scaled by each interested sky area. The best-fit parameters of the excess component are shown in Table~\ref{tab:se}. We also estimated the electron density of the excess component ($n_{\rm excess}$) assuming that it is uniform over a line-of-sight extent of $L_{\rm excess} =150$\,kpc, which is similar to the projected size of the substructure on the sky. For different values of $L_{\rm excess}$, the electron density scales as $L_{\rm excess}^{-1/2}$. In addition, we calculated the thermal pressure of the excess electrons ($p_{\rm excess}$), i.e., $n_{\rm excess} \times kT_{\rm excess}$. In above analyses, we fixed the metal abundance of the ICM at 0.38\,solar based on previous X-ray measurements \citep{Ota08}. 

\begin{figure*}
 \begin{center}
  \includegraphics[width=14cm]{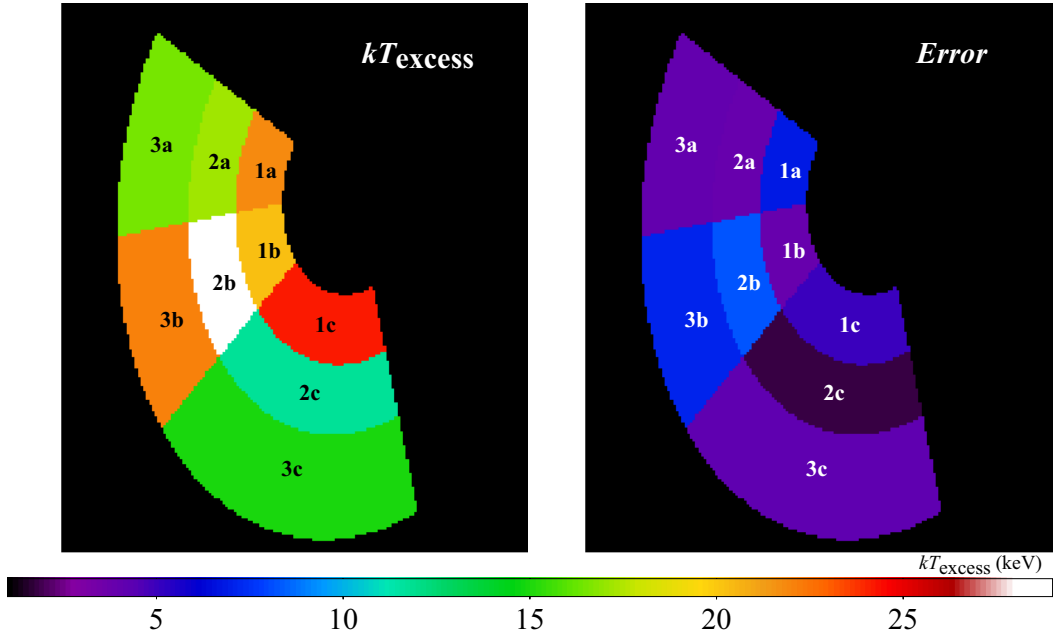}
\caption{
Temperature ({\it left}) and average of its statistical error ({\it right}) maps of the excess component ($kT_{\rm excess}$) in nine regions in units of keV. All the best-fit values are the same as shown in Table~\ref{tab:se}, while the error values are the average of upper and lower statistical errors of each region.
}
\end{center}
\label{fig:se_para_kT}
\end{figure*}

\begin{figure*}
 \begin{center}
  \includegraphics[width=14cm]{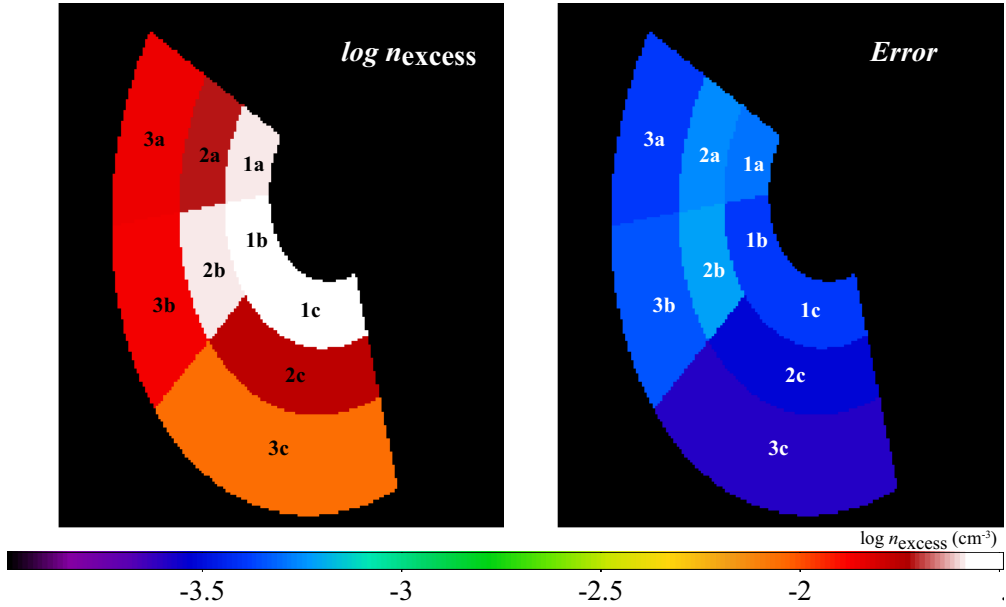} 
\caption{
Same as Figure~\ref{fig:se_para_kT} but for the logarithm of electron density of the excess component ($\log n_{\rm excess}$) in units of $\log$\,cm$^{-3}$. The values are approximately proportional to $(L_{\rm excess}/150\,{\rm kpc})^{-1/2}$.  
}
\end{center}
\label{fig:se_para_ne}
\end{figure*}

\begin{figure*}
 \begin{center}
  \includegraphics[width=14cm]{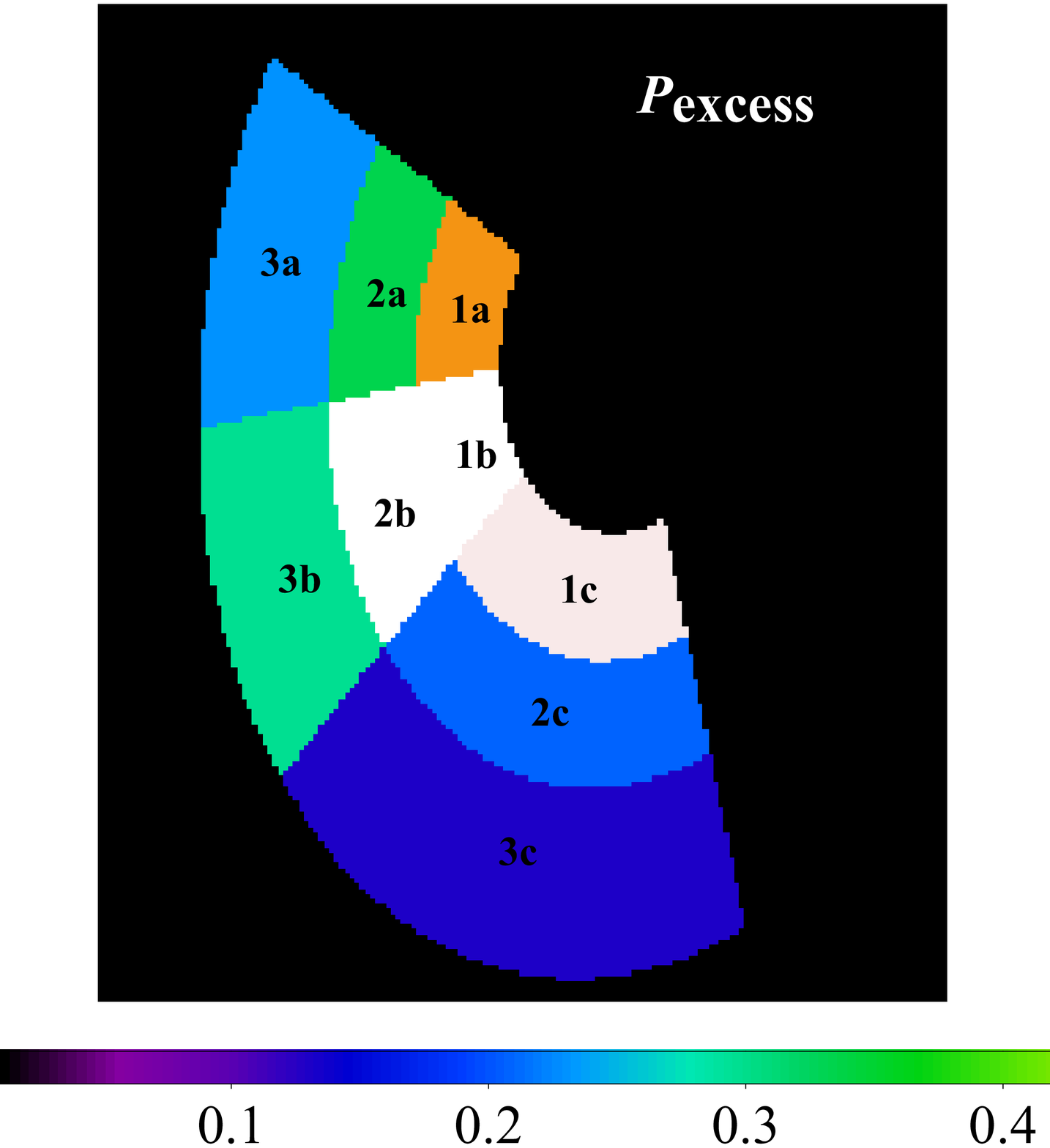} 
\caption{
Same as Figure~\ref{fig:se_para_kT} but for the thermal pressure of the excess electrons ($p_{\rm excess}$) in units of keV\,cm$^{-3}$. The values are also approximately proportional to $(L_{\rm excess}/150\,{\rm kpc})^{-1/2}$. 
}
\end{center}
\label{fig:se_para_p}
\end{figure*}

Figures~\ref{fig:se_para_kT}, \ref{fig:se_para_ne}, and \ref{fig:se_para_p} show the spatial maps of $kT_{\rm excess}, n_{\rm excess}$, and $p_{\rm excess}$ (left panels are their best-fits and right panels are corresponding statistical error), respectively. We find that the electron density of excess component in Region~1b is the highest among the nine regions, which agrees with the peak position in the X-ray residual image. On the other hand, while the statistical error is large, the temperature of the excess component in Region~2b is estimated to be $\sim 30$\,keV, which is the highest among the nine regions. The highest temperature region is located at $\sim 27''$ away from the center, which is well outside the cool core. Note that the X-ray emission in Region~2b is faint but the SZE signal is prominent. This means that the ALMA SZE data play a crucial role in identifying such excess hot gas. The thermal pressure of the excess electrons ($p_{\rm excess}$) is nearly constant among Regions~1b, 2b, and 1c, as presented in Table~\ref{tab:se}. This is in good agreement with the excess SZE signal shown in the right panel of Figure~\ref{fig:sbellip_9reg}. The thermal energy of the excess electrons in Region~1b, 2b, and 1c is estimated to be $\sim 1.1 \times 10^{61}\,h^{-2}$\,erg, $\sim 1.6 \times 10^{61}\,h^{-2}$\,erg, and $\sim 2.0 \times 10^{61}\,h^{-2}$\,erg, respectively. 

We estimated the Mach number $\cal M$ from the Rankine-Hugoniot jump condition, assuming the ratio of specific heats as $\gamma = 5/3$,
\begin{equation}
\frac{T_2}{T_1} = \frac{5{\cal M}^4 + 14{\cal M}^2 - 3}{16{\cal M}^2},
\end{equation}
where $T_1$ is the pre-shock temperature and $T_2$ is the post-shock temperature. Using the observed temperature jump between Region~2a and 2b and propagating the statistical errors, we obtain ${\cal M} = 1.68^{+0.69}_{-0.47}$. The adiabatic sound speed of the gas in Region~2a is estimated to be $2140^{+300}_{-180}$\,km\,s$^{-1}$, implying a shock speed of the gas in Region~2b as $3590^{+1560}_{-1050}$\,km\,s$^{-1}$.

\begin{table*}
\caption{Best-fit parameters to the X-ray spectra in each region shown in Figure \ref{fig:sbellip_9reg}. The metal abundance is fixed at each of 0.38\,$Z_{\odot}$. The photon counts are background-subtracted. $kT_{\rm single}$ is obtained by a single component model in the 12 regions. $kT_{\rm excess}$ and $n_{\rm excess}$ are obtained by a two-component fit in each region, fixing the temperature of the ambient component at the best-fit values for the same annulus region
(i.e., $kT_{\rm single}$ of Region~1d, 2d, and 3d). We show the statistical errors alone.
}
\label{tab:se}
  \begin{center}
    \begin{tabular}{ccccc}
      \hline \hline
  Region                            & 1a & 1b & 1c & 1d \\ \hline
 Photon counts in $0.4 - 7.0$\,keV & 4566 & 8023 & 10917 & 21083 \\
 $kT_{\rm single}$ (keV)   & $18.5^{+2.2}_{-2.1}$ & $18.2^{+1.9}_{-1.6}$ & $19.5^{+1.6}_{-1.6}$ & $15.7^{+0.7}_{-0.7}$ \\
 $kT_{\rm excess}$ (keV) & $21.8^{+8.6}_{-4.9}$ & $20.3^{+4.7}_{-3.1}$ & $24.2^{+6.1}_{-3.8}$ & -- \\
 $n_{\rm excess}$ ($10^{-2}$\,cm$^{-3}$\,$(L_{\rm excess}/150\,{\rm kpc})^{-1/2}$) & $2.53^{+0.06}_{-0.04}$ & $3.71^{+0.04}_{-0.04}$ & $2.97^{+0.04}_{-0.04}$ & -- \\
 $p_{\rm excess}$  (keV\,cm$^{-3}$\,$(L_{\rm excess}/150\,{\rm kpc})^{-1/2}$) & $0.552^{+0.218}_{-0.124}$ & $0.753^{+0.175}_{-0.115}$ & $0.719^{+0.181}_{-0.113}$ & -- \\
 \\ \hline
   Region                            & 2a & 2b & 2c & 2d \\ \hline
 Photon counts in $0.4 - 7.0$\,keV & 3208 & 4824  & 5996 & 12623 \\
 $kT_{\rm single}$ (keV)   & $17.8^{+3.1}_{-1.8}$ & $24.6^{+4.3}_{-3.6}$ & $14.8^{+1.5}_{-1.5}$ & $19.5^{+1.6}_{-1.6}$ \\
 $kT_{\rm excess}$ (keV) & $17.2^{+4.8}_{-2.9}$ & $29.1^{+9.4}_{-6.9}$ & $11.7^{+2.0}_{-1.5}$ & -- \\
 $n_{\rm excess}$ ($10^{-2}$\,cm$^{-3}$\,$(L_{\rm excess}/150\,{\rm kpc})^{-1/2}$) & $1.93^{+0.06}_{-0.05}$ & $2.55^{+0.06}_{-0.06}$ & $1.77^{+0.03}_{-0.03}$ & -- \\
 $p_{\rm excess}$  (keV\,cm$^{-3}$\,$(L_{\rm excess}/150\,{\rm kpc})^{-1/2}$) & $0.332^{+0.093 }_{-0.057}$ & $0.742^{+0.240}_{-0.177}$ & $0.207^{+0.036 }_{-0.027}$ & -- \\
 \\ \hline
   Region                            & 3a & 3b & 3c & 3d \\ \hline
 Photon counts in $0.4 - 7.0$\,keV & 3337 & 3682  & 4281 & 9987 \\
 $kT_{\rm single}$ (keV)   & $17.9^{+3.1}_{-2.5}$ & $21.3^{+4.0}_{-3.1}$ & $17.2^{+2.2}_{-1.1}$ & $20.2^{+1.9}_{-1.7}$ \\
 $kT_{\rm excess}$ (keV) & $16.5^{+4.9}_{-3.2}$ & $22.1^{+8.9}_{-5.2}$ & $14.9^{+4.8}_{-3.6}$ & -- \\
 $n_{\rm excess}$ ($10^{-2}$\,cm$^{-3}$\,$(L_{\rm excess}/150\,{\rm kpc})^{-1/2}$) & $1.40^{+0.04}_{-0.04}$ & $1.34^{+0.05}_{-0.04}$ & $0.88^{+0.03}_{-0.02}$ & -- \\
 $p_{\rm excess}$  (keV\,cm$^{-3}$\,$(L_{\rm excess}/150\,{\rm kpc})^{-1/2}$) & $0.231^{+0.069}_{-0.045}$ & $0.296^{+0.120}_{-0.070}$ & $0.131^{+0.042 }_{-0.032 }$ & -- \\ 
      \hline \hline
    \end{tabular}
  \end{center}
\end{table*}

\subsection{Combined analysis of SZE and X-ray imaging data for the excess component}
\label{sec:se_image}

As an alternative measure of the gas temperature and electron density, we combined the SZE and X-ray images. We used {\tt SPEX} version 3.04.00 for the X-ray data analysis and {\tt CASA} version 5.1.1 for the SZE data analysis. Note that the analysis in this section does not rely on X-ray spectra and is fully independent of that presented in Section~\ref{sec:X-ray_spec}.

\begin{figure*}
 \begin{center}
  \includegraphics[width=14cm]{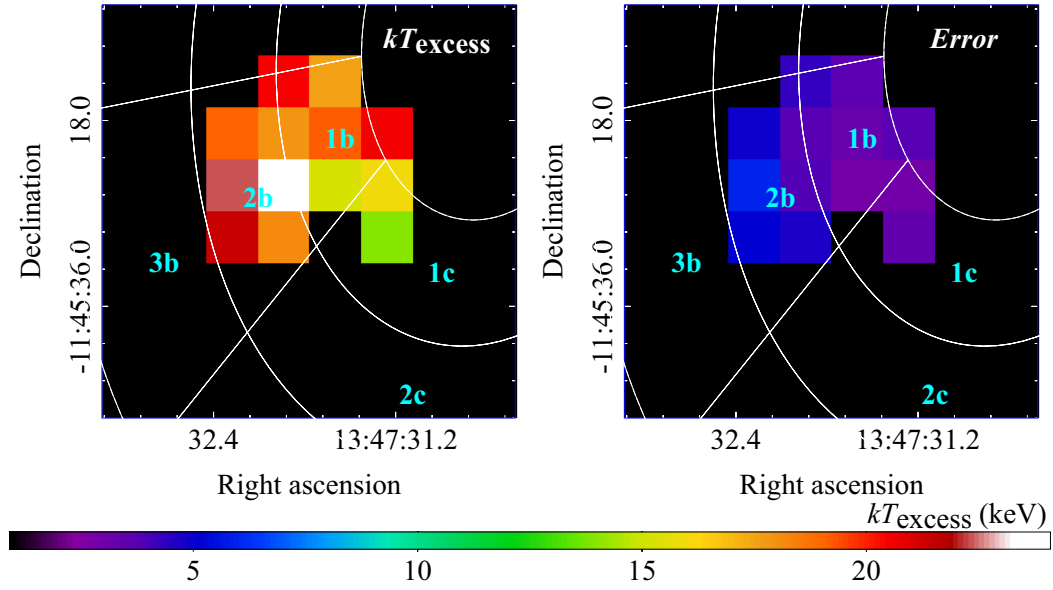} 
 \end{center}
\caption{Temperature of the excess component in units of keV ({\it left}) and its statistical error ({\it right}) inferred from the ALMA SZE and \Chandra ~X-ray images, assuming a constant line-of-sight length of $L_{\rm excess}=150$\,kpc. The values are approximately proportional to $(L_{\rm excess}/150\,{\rm kpc})^{-1/2}$. Regions for which S/N of either data are less than $4\sigma$ are left blank (black colored). For reference, the elliptical annuli shown in Figure~\ref{fig:sbellip_9reg} are overlaid. Note that X-ray spectra are not used in computing the quantities plotted in this figure.
}
\label{fig:se_kt}
\end{figure*}

\begin{figure*}
 \begin{center}
  \includegraphics[width=14cm]{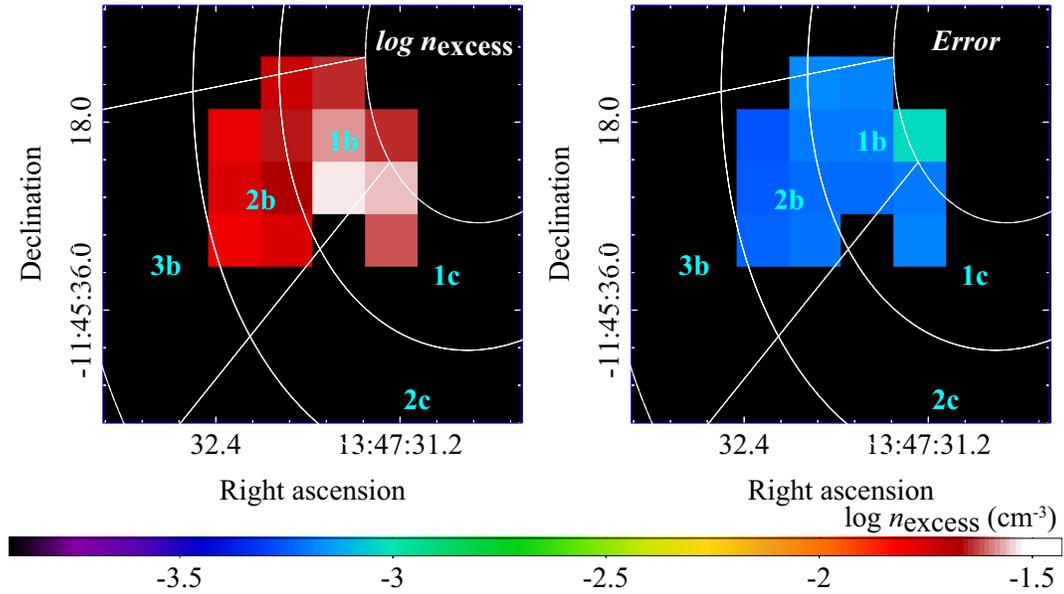} 
 \end{center}
\caption{Same as Figure~\ref{fig:se_kt}, but for the logarithm of electron density ({\it left}) and its statistical error ({\it right}) in units of cm$^{-3}$. The values are also approximately proportional to $(L_{\rm excess}/150\,{\rm kpc})^{-1/2}$.
}
\label{fig:se_ne}
\end{figure*}

We measured the intensities of X-ray ($\Delta I_{\rm X}$) and SZE ($\Delta I_{\rm SZ}$) of the X-ray and SZE residual images shown in Figure~\ref{fig:sbellip_xsz}. Both images were binned so that the angular resolutions of \Chandra ~and ALMA have a common pixel size of $5'' \times 5''$. The two intensities are then described as 
\begin{align}
\Delta I_{\rm X}   &\propto n_{\rm excess}^2 \times \Lambda(T_{\rm excess}) \times L_{\rm excess}, \label{eq:X} \\
\Delta I_{\rm SZ} &\propto n_{\rm excess} \times kT_{\rm excess} \times L_{\rm excess} \times f_{r}(T_{\rm excess}) \times f_{c}, \label{eq:SZ}
\end{align}
respectively, where $\Lambda$ is the X-ray emissivity over the energy range $0.4-7.0$\,keV, $f_r$ is the relativistic correction to the SZE intensity \citep{Itoh04}, and $f_c$ is a correction factor for the missing flux in the ALMA data. We set $f_c$ to be equal to the parameter $c_1=0.88$ in Equation (2) of \cite{Kitayama16} derived from detailed imaging simulations for \rxj \footnote{The bulk of the missing flux in the ALMA data resides in a constant offset in the brightness, represented by another parameter $c_0$ in Equation (2) of \cite{Kitayama16}. Such an offset, however, is fully removed in the residual image and does not affect the analysis of the present paper.}. We then solved Equations~(\ref{eq:X}) and (\ref{eq:SZ}) for $n_{\rm excess}$ and $kT_{\rm excess}$, respectively, assuming $L_{\rm excess} = 150$\,kpc as in Section~\ref{sec:X-ray_spec}. For different values of $L_{\rm excess}$, they approximately scale as $kT_{\rm excess} \propto L_{\rm excess}^{-1/2}$ and $n_{\rm excess} \propto L_{\rm excess}^{-1/2}$, if weak dependences of $\Lambda$ and $f_r$ on temperature is neglected. We focused on the high significance pixels for which the signal-to-noise ratios (S/N) of $\Delta I_{\rm X}$ and $\Delta I_{\rm SZ}$ are both over $4\sigma$. Figure~\ref{fig:se_kt} shows the measured temperature of the excess component ({\it left}) and its statistical error ({\it right}). Figure~\ref{fig:se_ne} is the same as Figure~\ref{fig:se_kt} but for the electron density of the excess component. The typical errors of temperature and electron density are about 4\,keV and $0.05 \times 10^{-2}$\,cm$^{-3}$, respectively.

We find that the highest temperature region is located in Region~2b. This result is consistent with that measured with an independent method using the X-ray spectra (Section~\ref{sec:X-ray_spec}). We also find that the highest electron density region is in Region~1b. In addition, we estimate the thermal energy of the excess electrons to be $(1.9 \pm 0.1\,({\rm stat.}) \pm 0.7\,({\rm sys.})) \times 10^{61}$\,$h^{-2}$\,erg from the SZE residual image directly, within a radius of 50\,$h^{-1}$\,kpc around the peak of the SZE residual signal. Further details of the systematic uncertainty are mentioned in Section~\ref{sec:sys}. The angular resolution and sensitivity of the previous SZE observations did not match those of the X-ray data. The ALMA SZE data allow us, for the first time, the detailed comparison with the X-ray data with comparable resolution and sensitivity.

\subsection{Nature of gas sloshing in the core}
\label{sec:sloshing}

The spatial scale of the dipolar pattern is estimated to be $40\,h^{-1}$\,kpc by applying the definition of the size described in \cite{Ueda17}.
We analyzed the X-ray spectra of the ICM in the region of the dipolar pattern to examine its thermodynamic property. 
The temperature toward the positive and the negative excess is estimated as $8.8^{+0.2}_{-0.2}$\,keV and $10.6^{+0.4}_{-0.3}$\,keV, respectively.
The metal abundances are $0.47^{+0.03}_{-0.03}$\,\ZO ~for the positive excess and $0.37^{+0.05}_{-0.05}$\,\ZO ~for the negative excess.
These properties are as expected for gas sloshing, i.e., metal rich, cool, and dense gas is moving around the central galaxy.
On the other hand, we find the absence of excess SZE signal in the core, which provides the first direct evidence that the disturbed gas is nearly in pressure equilibrium. This indicates that the gas sloshing motion is sub-sonic. 

Here, we shall introduce the ``equation of state'' of the perturbation in gas, $w \equiv \Delta p / \Delta \rho$. For similar analyses, see \cite{Churazov16}, \cite{Khatri16}, and references therein. If the perturbation is adiabatic, then $w$ is equal to the sound speed squared, i.e., $w=c_s^2$. If the perturbation is isobaric, we should find $w \ll c_s^2$. The SZE residual image gives an estimate of the pressure perturbation $\Delta p$, whereas the X-ray residual image gives an estimate of the density perturbation $\Delta \rho$ as
\begin{equation}
\frac{|\Delta I_{\rm X}|}{\langle I_{\rm X} \rangle} \simeq \frac{\Delta \langle n^2 \rangle}{\langle n^2 \rangle} \simeq \frac{2 \Delta \rho}{\sqrt{\langle \rho^2 \rangle}} \label{eq:cs}
\end{equation}
because $|\Delta I_{\rm X}| \ll \langle I_{\rm X} \rangle$, where $\langle I_{\rm X} \rangle$ is the mean X-ray surface brightness, $\langle n^2 \rangle$ is the mean square electron number density, and $\langle \rho^2 \rangle$ is the mean square gas mass density. First, to cover the dipolar pattern in the X-ray residual image, we used the ellipse with the same geometry as mentioned in Section~\ref{sec:ximage}, with a semi-major axis length of $22''$, excluding the southeast quadrant. We extracted the X-ray spectrum for this elliptical sector and carried out the X-ray spectral analysis. Assuming the line-of-sight depth of 100\,kpc, the mean ICM temperature and the mean electron number density, $\sqrt{\langle n^2 \rangle}$, in this region are $9.8 \pm 0.2$\,keV and $(8.24 \pm 0.02) \times 10^{-2}$\,cm$^{-3}$, respectively. Next, using the X-ray and SZE residual images and assuming the line-of-sight depth of 100\,kpc, we measured $\Delta \rho$, $\Delta p$, and their statistical errors within the same elliptical sector. We then find that the velocity inferred from the equation of state is $\sqrt{w} = 420^{+310}_{-420}$\,km\,s$^{-1}$. This inferred velocity is much lower than the adiabatic sound speed of the 10\,keV ICM, $\sim 1630$\,km\,s$^{-1}$. The motion of gas sloshing is therefore consistent with being sub-sonic and isobaric.

\subsection{Systematic uncertainty of the ICM properties}
\label{sec:sys}

The systematic uncertainty of the ICM temperature measured with the ACIS is estimated to be $\sim 20$\,\% for galaxy clusters with high temperature (over 10\,keV) ICM like \rxj ~\citep[e.g.,][]{Reese10, Nevalainen10, Schellenberger15}. Note that the statistical uncertainty of the temperature of the excess component is $20 \sim 30$\,\% in the nine regions. We evaluated the impact of the systematic uncertainty on the X-ray and the SZE data for the combined analysis. We applied the 4\,\% uncertainty of the ACIS effective area \footnote{\url{http://cxc.harvard.edu/cal/summary/Calibration_Status_Report.html}}, the 6\,\% uncertainty of the ALMA flux calibration \citep[see Section~2 of][]{Kitayama16}, and 1.3\,$\mu$Jy\,arcsec$^{-2}$ for the ALMA missing flux correction \citep[see Section~4.2 of][]{Kitayama16} to the data. The error of the ICM temperature and density including the systematic uncertainty is about twice larger than that shown in the right panels of Figure~\ref{fig:se_kt} and \ref{fig:se_ne}; for example, the error for the highest temperature region (23.2\,keV) changes from $\pm 3.9$\,keV to $\pm 8.1$\,keV. Note that these systematic uncertainties are only relevant to the absolute values of derived parameters but not their relative trends, in which we are mainly interested.

\section{\HST ~strong-lensing analysis}
\label{sec:hst}

\subsection{Central mass distribution}
\label{sec:mass}

To reveal the geometry of mergers in \rxj, we analyzed the central mass distribution derived by SL. We used the {\tt GLAFIC} software package \citep{Oguri10} for our mass modeling. {\tt GLAFIC} adopts a parametrized mass model and derives the best-fit mass map that reproduces the observed positions of SL multiple images. The software has been used for SL mass modeling of many massive clusters \citep[e.g.,][]{Oguri12, Kawamata16} and has also been shown to recover central mass distributions of simulated clusters remarkably well \citep{Meneghetti17}.

We followed the standard procedure to model massive clusters in which we adopted a few dark halo components, cluster member galaxies, and the external perturbations on the lens potential \citep[see e.g.,][]{Kawamata16}. For dark halo components, we assumed an elliptical extension of the Navarro-Frenk-White (NFW) model \citep{Navarro97}. Each dark halo component is characterized by a set of parameters including the virial mass, the concentration parameter, positions, the eccentricity of ellipse, and its position angle as model parameters. For member galaxies, we used the scaling relations to reduce the number of model parameters. We constructed a cluster member galaxy catalog by selecting galaxies near the red-sequence in the \HST ~F475W-F814W color-magnitude diagram, where the magnitudes are {MAG\_APER} of {\tt SExtractor} \citep{Bertin96} with $1''$ diameter aperture. The mass distribution of the member galaxies was assumed to follow pseudo-Jaffe ellipsoids whose velocity dispersions and truncation radii scale with galaxy luminosity $L_{\rm gal}$ as $\sigma\propto L_{\rm gal}^{1/4}$ and $r_{\rm trunc}\propto L_{\rm gal}^{1/2}$, respectively. The normalizations of these two scaling relations were treated as model parameters. In addition, to improve the fit, we included external perturbations described by a multipole Taylor expansion of the form $\phi \propto r^2\cos m(\theta-\theta_*)$. We included terms with $m=2$ (external shear), $3$, and $4$. Each term is specified by its amplitude and position angle as model parameters. 

For observational constraints, we used the positions of 21 multiple images from six multiple image sets of the lensed galaxies. These multiple images are identified in previous work on this cluster \citep[][]{Halkola08, Kohlinger14, Zitrin15}. In this paper, we reexamined the validity of the previous multiple image identifications in the course of our own mass modeling, and adopted only secure multiple images for our analysis. The redshift of one multiple image set was fixed as the spectroscopic value, whereas for those of four multiple image sets, we included Gaussian priors based on their photometric redshifts. We assumed a positional error of $0\farcs 5$ for all the multiple images, which is a standard value in SL mass modeling \citep[e.g.,][]{Kawamata16}. Table~\ref{tab:SLdata} summarizes the data of multiple images for our SL analyses.

\begin{table*}
\caption{Data of 21 multiple images from six multiple image sets of the lensed galaxies used for our SL analyses.}
\label{tab:SLdata}
  \begin{center}
    \begin{tabular}{ccccl}
      \hline \hline
ID  &  R.A. (deg)     & Dec. (deg)      & $z_{\rm obs}$\tablenotemark{a} & Reference\tablenotemark{b} \\ \hline
 1.1 & 206.887550 & -11.757544 & $2.2\pm0.3$ &  1 \\
 1.2 & 206.878783 & -11.753378 & &  \\
 1.3 & 206.877675 & -11.759375 & &  \\
 1.4 & 206.885183 & -11.748411 & &  \\
 1.5 & 206.869604 & -11.747325 & &  \\
 2.1 & 206.882608 & -11.764414 & $1.75$      &  1, 2 \\
 2.2 & 206.871675 & -11.760722 & &  \\
 2.3 & 206.872312 & -11.761255 & &  \\
 3.1 & 206.874179 & -11.745133 & $4.1\pm0.3$ &  1, 2 \\
 3.2 & 206.872500 & -11.746467 & &  \\
 3.3 & 206.871162 & -11.748361 & &  \\
 3.4 & 206.872271 & -11.757350 & &  \\
 3.5 & 206.891250 & -11.760539 & &  \\
 3.6 & 206.876079 & -11.744372 & &  \\
 4.1 & 206.884138 & -11.741794 & $3.6\pm0.3$ &  1 \\
 4.2 & 206.882975 & -11.741230 & &  \\
 4.3 & 206.892412 & -11.751589 & &  \\
 5.1 & 206.878429 & -11.749203 & NA    &  1 \\
 5.2 & 206.878317 & -11.749703 & &  \\
 6.1 & 206.882320 & -11.742199 & $6.6\pm0.3$  &  3 \\
 6.2 & 206.875980 & -11.741229 & &  \\
\hline \hline
    \end{tabular}
  \end{center}
\tablenotetext{a}{$z_{\rm obs}$ without an error bar indicates a spectroscopic redshift which is fixed during mass modeling, whereas those with error bars are photometric redshifts. In the mass modeling, we regarded source redshifts of the five multiple image sets without the spectroscopic redshifts as model parameters, but included Gaussian priors for those with photometric redshifts.
}
\tablenotetext{b}{References: 1 -- \cite{Halkola08} \citep[see also][]{Bradac08}; 2 -- \cite{Kohlinger14}; 3 -- \cite{Zitrin15}.}
\end{table*}

Our fiducial model contains two dark halo components, which are supposed to model mass distributions near the brightest cluster galaxy (BCG) of the main cluster and the second BCG (hereafter 2nd BCG) that is considered to be the BCG of the subcluster, respectively. This model contains 37 model parameters, which are constrained by 46 observational data points. We also considered another model which contains three dark halo components, as such model may reveal a mass component which does not correspond to either the BCG or the 2nd BCG. In this model, the number of model parameters increases to 43. We evaluated statistical errors of model parameters, such as masses of dark halo, concentration parameters, and mass peaks with Markov chain Monte Carlo (MCMC) methods.

\begin{figure*}
 \begin{center}
  \includegraphics[width=8.5cm]{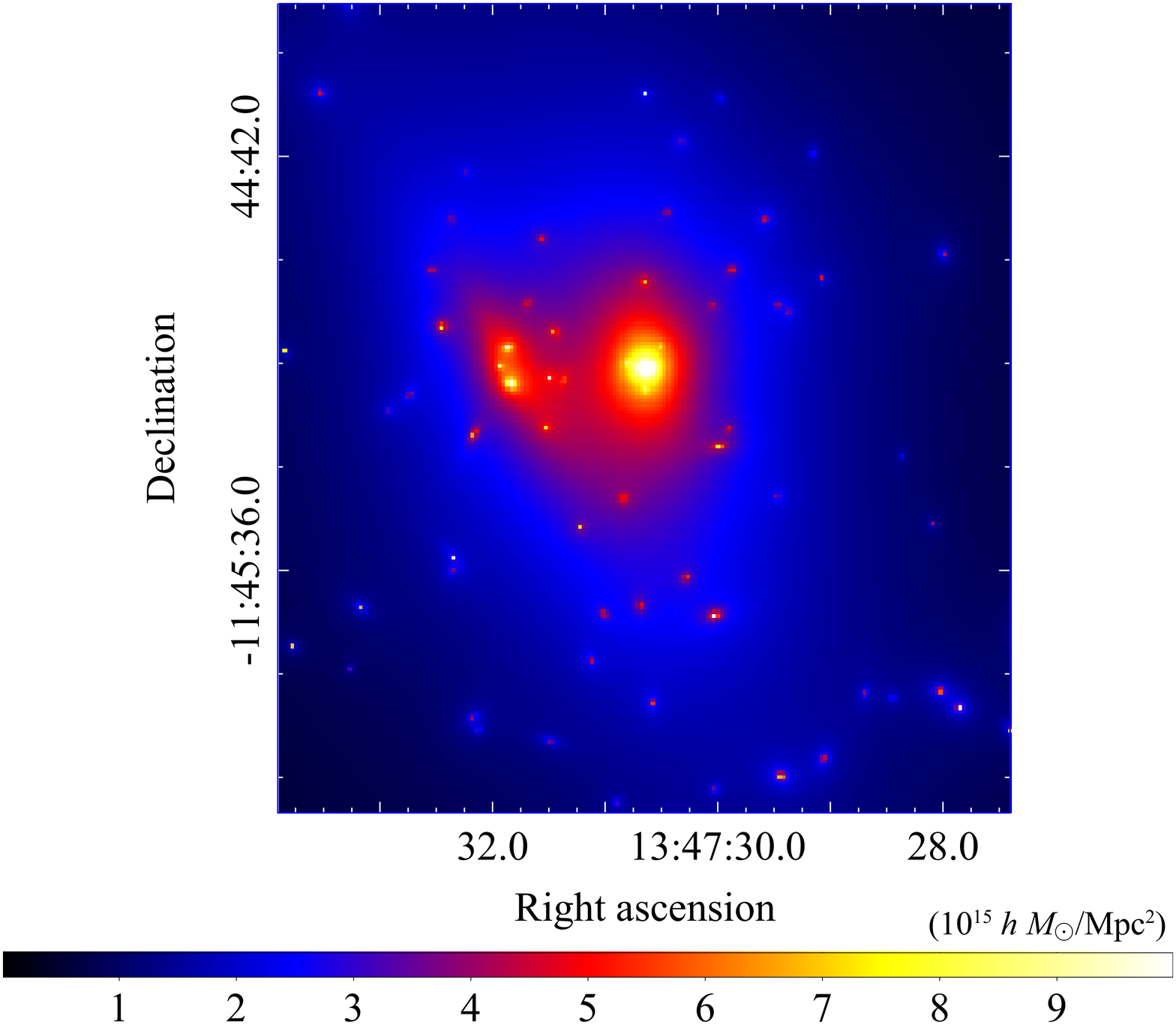}
  \includegraphics[width=8.5cm]{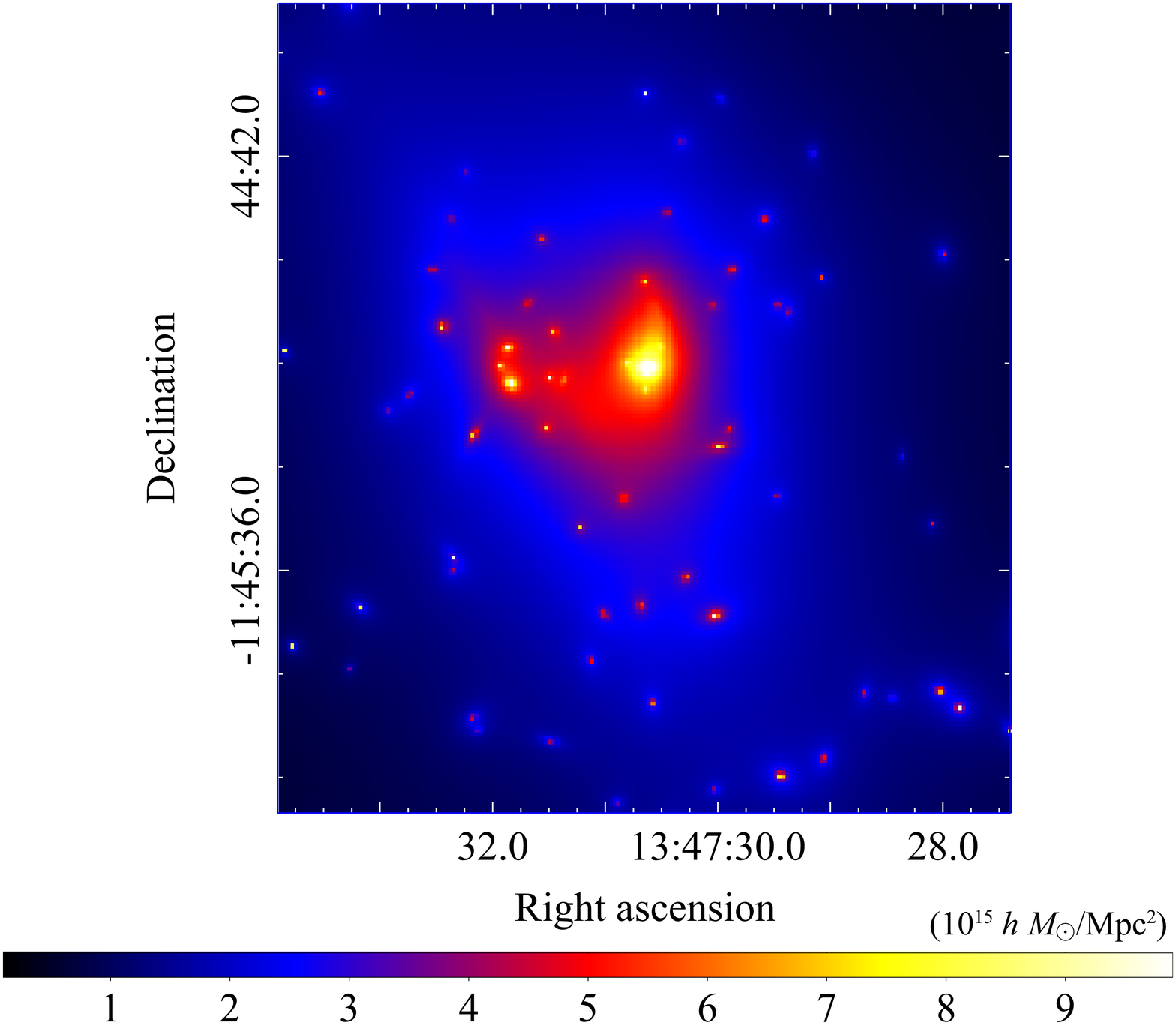} 
 \end{center}
\caption{
Surface mass density map of \rxj ~measured with SL assuming two dark halo components ({\it left}) or three ones ({\it right}) in units of $10^{15}$\,$h$\,\MO\,Mpc$^{-2}$. The entire morphology is nearly unchanged by an addition of the third dark matter component.
}
\label{fig:SL}
\end{figure*}

The surface mass density map obtained from the two dark halo components is shown in the left panel of Figure~\ref{fig:SL}. The best-fit parameters are summarized in Table~\ref{tab:SL}. The locations of the two mass peaks are treated as free parameters, while we find that the two mass peaks are in good agreement with the positions of the BCG and the 2nd BCG. The estimated virial masses are $7.2^{+2.1}_{-1.9} \times 10^{14}$\,$h^{-1}$\MO ~for the main cluster and $2.8^{+2.4}_{-1.1} \times 10^{14}$\,$h^{-1}$\MO ~for the subcluster. Their concentration parameters are estimated to be $c = 6.6^{+1.0}_{-0.9}$ and $c = 4.8^{+2.2}_{-1.6}$, respectively. The obtained mass and concentration parameter are correlated strongly, therefore the systematic uncertainty of the mass estimates might be large. Nevertheless their mass ratio indicates that \rxj ~is a major merging cluster. The total mass of the subcluster has been previously estimated using the excess X-ray flux and applying the luminosity - mass relation to be $(3.4 \pm 1.7) \times 10^{14}$\,$h^{-1}$\,\MO ~\citep{Johnson12} and $\sim 2.3 \times 10^{14}$\,$h^{-1}$\,\MO ~\citep{Kreisch16}, which are consistent with that by our SL analysis\footnote{Both masses were originally estimated based on the Hubble constant of 70\,km\,s$^{-1}$\,Mpc$^{-1}$, while we re-calculated them using our $h$ to adjust to this paper.}. The entire mass distribution of \rxj ~is slightly elongated along the northeast direction.

The right panel of Figure~\ref{fig:SL} shows the surface mass density map obtained from the three dark halo components. The mass of the third component is weakly constrained to be $1.4 \times 10^{14}$\,$h^{-1}$\,\MO ~(95\% CL). The location of the third component is also poorly determined, but it lies to the northeast of the BCG. These indicate that the mass distribution of dark matter (DM) near the center and in the southeast quadrant of \rxj ~is well modeled by the two dark halo components. We therefore neglect the third dark halo component in the rest of this paper.

\begin{table*}
\caption{Best-fit parameters of mass distributions of the main cluster and the subcluster measured by SL}
\label{tab:SL}
  \begin{center}
    \begin{tabular}{ccc}
          \hline \hline
          	& Main cluster & Subcluster \\ \hline
	Mass ($10^{14}\,h^{-1}$\MO) & $7.2^{+2.1}_{-1.9}$ & $2.8^{+2.4}_{-1.1}$  \\
	Concentration parameter & $6.6^{+1.0}_{-0.9}$ & $4.8^{+2.2}_{-1.6}$ \\
	Offset of mass peak relative to each BCG (arcsec) \tablenotemark{a} & ($0.2^{+0.8}_{-0.8}$, $0.1^{+1.0}_{-0.8}$) & ($0.5^{+1.8}_{-2.3}$, $3.7^{+2.9}_{-3.9}$) \\
	Semi-minor to semi-major axis ratio & $0.62^{+0.04}_{-0.05}$ & $0.31^{+0.06}_{-0.06}$ \\
	Position angle (deg) & $1.1^{+4.6}_{-5.0}$ & $31.7^{+2.0}_{-2.2}$ \\
          \hline \hline	
    \end{tabular}
  \end{center}
\tablenotetext{a}{The positions of the BCG and the 2nd BCG are (RA, Dec.) = (13h47m30.639s, -11d45m09.589s) and (RA, Dec.) = (13h47m31.870s, -11d45m11.20s), respectively.}
\end{table*}

\subsection{Gas stripping and sloshing}
\label{sec:strip}

\begin{figure*}
 \begin{center}
  \includegraphics[width=8.5cm]{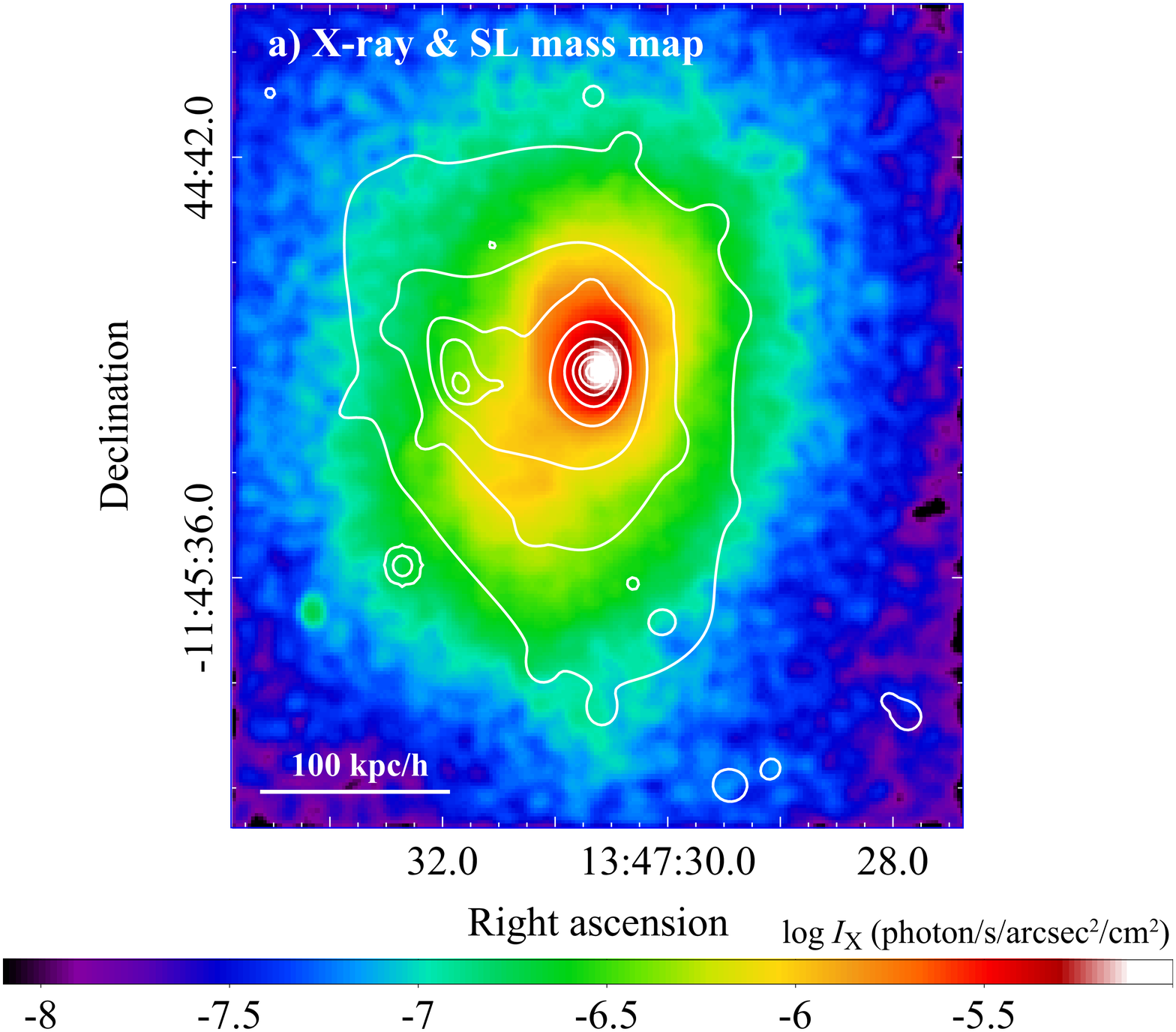} 
  \includegraphics[width=8.5cm]{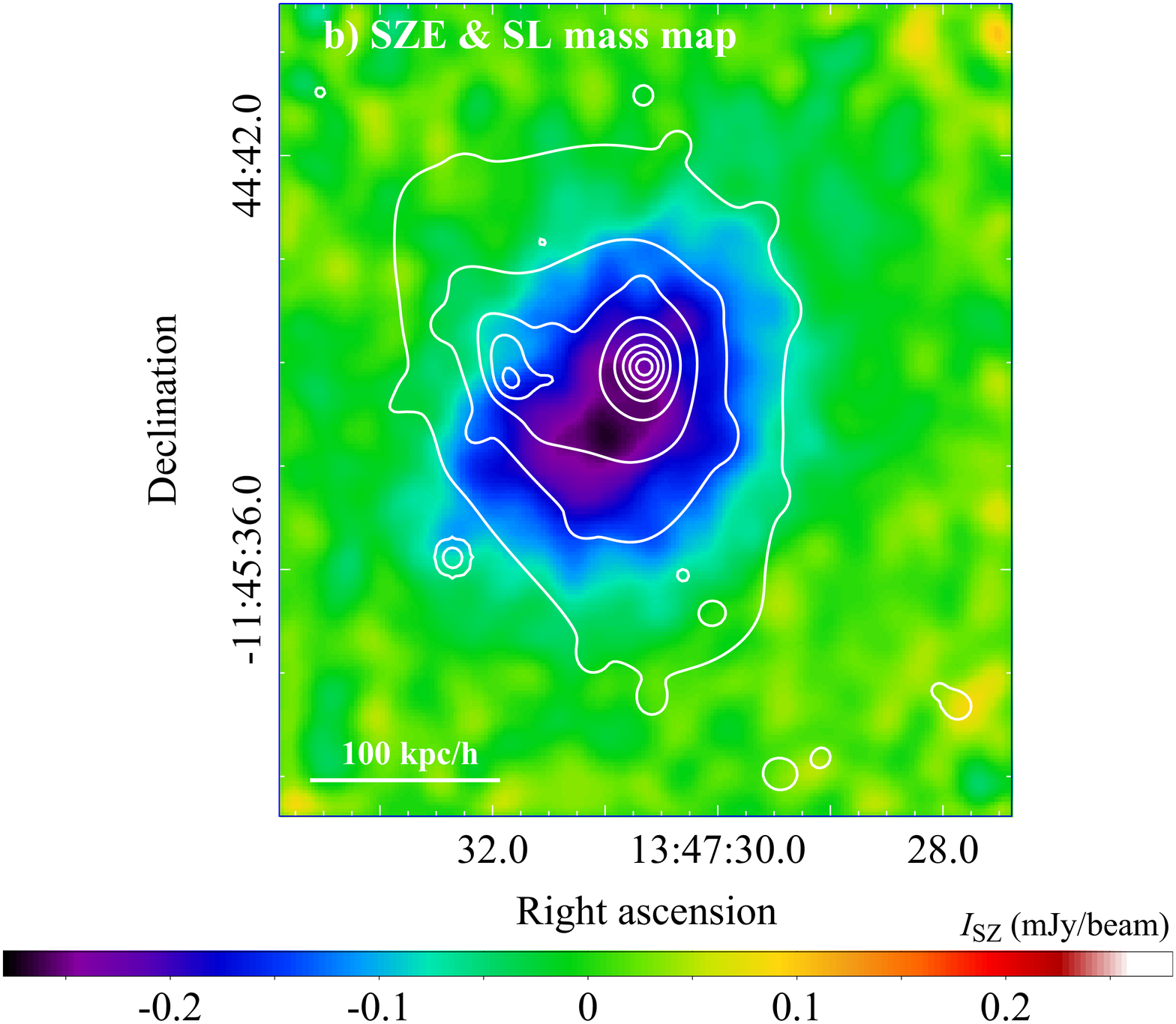} 
  \includegraphics[width=8.5cm]{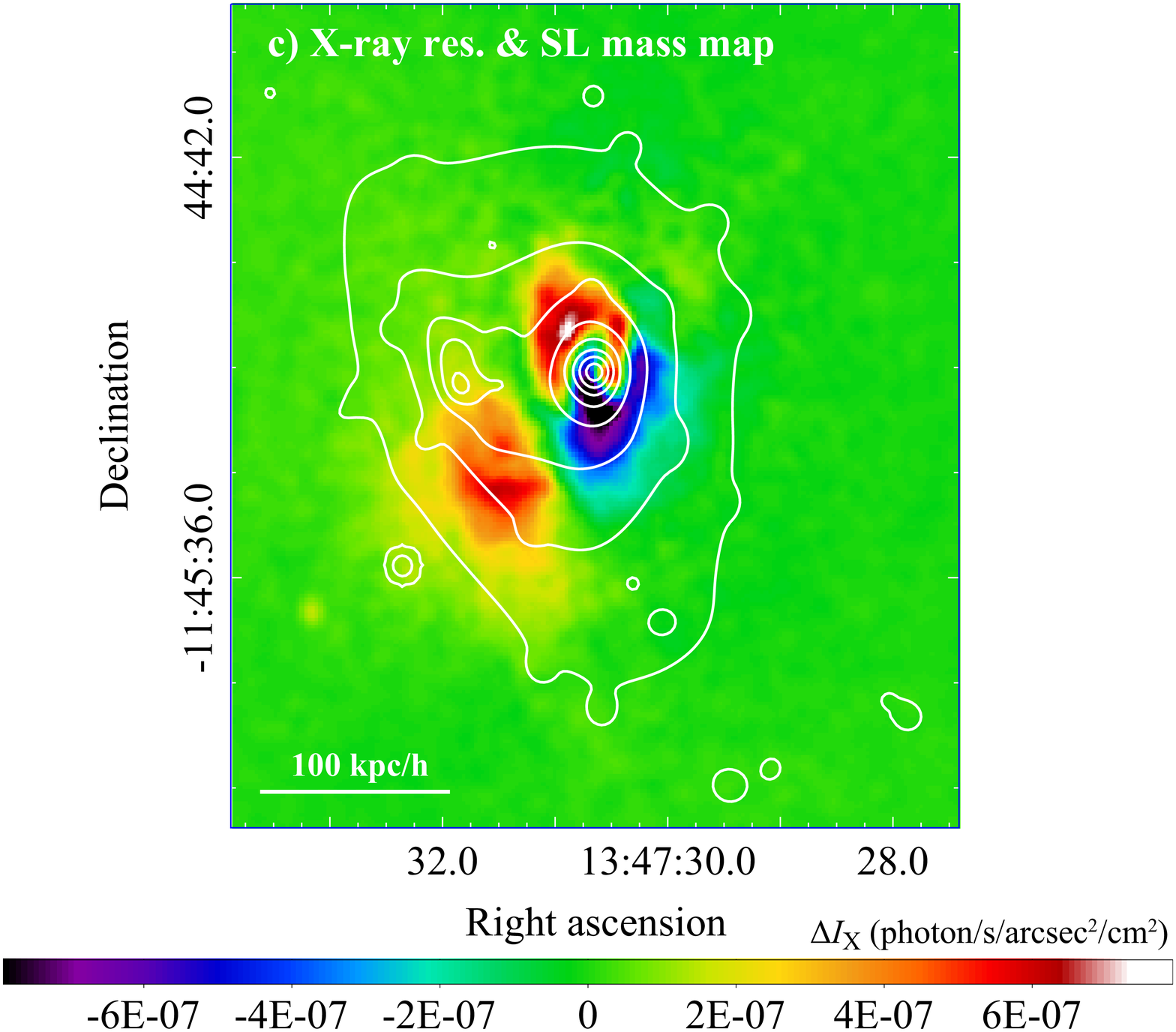}
  \includegraphics[width=8.5cm]{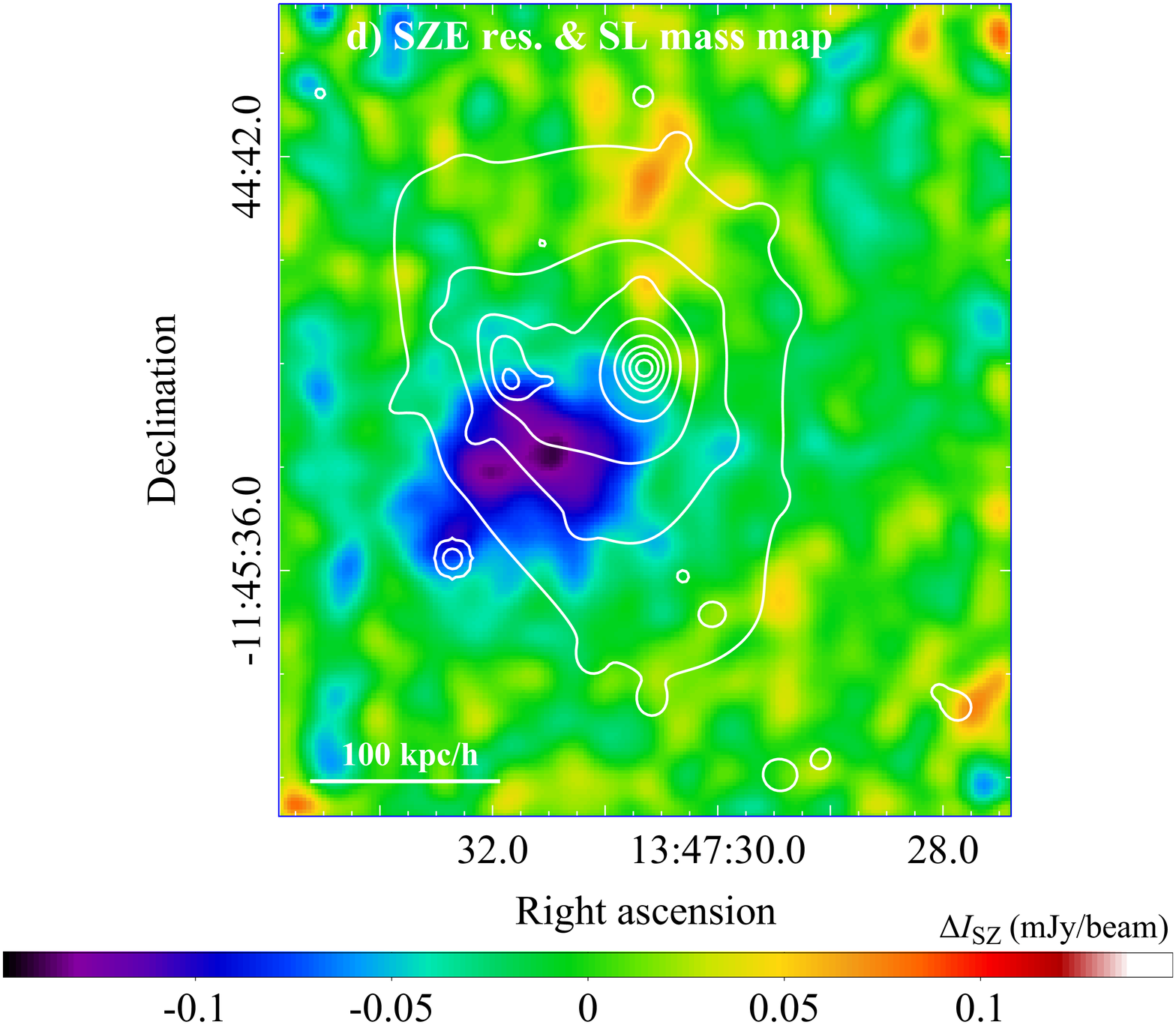}
 \end{center}
\caption{
Comparison of morphology between the surface mass density map derived from two dark halo components and other wavelength images. 
{\it Top left:} White contours of the surface mass density map are overlaid on the X-ray surface brightness image. The contours indicate levels of 2.0, 3.0, 4.0, $\cdots$, 9.0, $10.0 \times 10^{15}$\,$h$\,\MO\,Mpc$^{-2}$, respectively. This X-ray image is the same as the left panel of Figure~\ref{fig:sb}. The mass peak of the main cluster coincides with the X-ray centroid, while no X-ray counterparts are found at the mass peak of the subcluster. 
{\it Top right:} Same as the top left panel but for the SZE image. This SZE image is the same as the right panel of Figure~\ref{fig:sb}. No SZE counterparts are found at the mass peak of the subcluster. 
{\it Bottom left:} Same as the top left panel but for the X-ray residual image. This X-ray residual image is the same as the left panel of Figure~\ref{fig:sbellip_xsz}. The clear offset between the mass peak of the subcluster and the X-ray substructure is found. 
{\it Bottom right:} Same as the top left panel but for the SZE residual image. This SZE residual image is the same as the right panel of Figure~\ref{fig:sbellip_xsz}. As shown in this top right panel, the offset between the SZE substructure and the mass peak of the subcluster is found. 
}
\label{fig:X-SZE-SL}
\end{figure*}

We find a clear offset between the mass peak of the subcluster and the position of the substructure in X-rays and the SZE (the bottom panels of Figure~\ref{fig:X-SZE-SL}). The second mass peak is away from the peak position of the X-ray substructure with statistical significance of $5.5\sigma$. The second DM component is most likely associated with the infalling subcluster, suggesting that the gas that was originally in the subcluster is stripped by ram-pressure of the main cluster. 

The top left panel of Figure~\ref{fig:X-SZE-SL} shows that the X-ray centroid coincides with the mass peak of the main cluster even though the ICM in the cool core is sloshing. The mass peak of the main cluster is located at the interface between the positive and the negative excesses of the dipolar pattern (see the bottom left panel of Figure~\ref{fig:X-SZE-SL}). These features indicate that part of the gas in the cool core is moving around the BCG, but the DM distribution is not disturbed.

\section{Discussion}
\label{sec:discussion}

\subsection{Origin of the excess hot ICM}

The presence of very hot ICM in excess of 20\,keV is inferred in \rxj ~for the earlier SZE and X-ray observations \citep{Kitayama04, Ota08}. Combining the X-ray data of \Chandra ~with the SZE data of ALMA, we have confirmed the temperature of the excess hot ICM by two independent methods and have improved precision of its position. The excess hot ICM is located at a distance of $27''$ (or $\sim 109\,h^{-1}$\,kpc) to the southeast from the cluster center.

Following Section~\ref{sec:sloshing}, we also measure the equation of state of the excess hot ICM. We compute  $\Delta p$ and $\Delta \rho$ for an elliptical region with semi-major and semi-minor axis lengths of $14''$ and $12''$, respectively, and a position angle of $-70^{\circ}$ around the centroid of the excess SZE signal. We assume the line-of-sight depth of 150\,kpc. Given that $\Delta I_{\rm X} > \langle I_{\rm X} \rangle$ in this region, we adopt $n_{\rm excess}$ given in Equation~(\ref{eq:X}) as a better proxy for $\Delta \rho$ than Equation~(\ref{eq:cs}). On the other hand, $\Delta p$ is directly computed from the SZE residual signal, $\Delta I_{SZ}$. We analyze the X-ray spectrum in this region using the same procedure as described in Section~\ref{sec:X-ray_spec}. The electron number density and temperature of the excess hot ICM is $(3.29 \pm 0.02) \times 10^{-2}$\,cm$^{-3}$ and $20.2^{+2.1}_{-1.8}$\,keV, respectively. We then find that the velocity inferred from the equation of state of the excess hot ICM is $1970 \pm 150$\,km\,s$^{-1}$, which matches $\sim 85$\,\% of the adiabatic sound speed of the 20\,keV ICM, $\sim 2310$\,km\,s$^{-1}$. This supports picture that the pressure perturbation in the southeast quadrant is induced by shock.

The excess hot ICM in the southeast quadrant is most likely associated with a major merger. The result of SL analysis indicates that the mass peak of the subcluster is offset from the location of the southeast substructure. Our results imply that the gas that was originally in the subcluster is stripped by ram-pressure of the main cluster. As mentioned above, the pressure perturbation of the southeast substructure in the SZE residual image is consistent with that created by shock, so it is plausible that this stripped gas has been heated during the major merger. Details of shock-heating processes in \rxj ~using numerical simulations will be presented elsewhere.

\subsection{Major merger and sloshing cool core}

We have also found a dipolar pattern in the core of \rxj ~in its X-ray residual image. This is direct evidence that the core is experiencing gas sloshing, whose presence was suggested before by \cite{Johnson12} and \cite{Kreisch16}. We find that the ICM properties of the disturbed gas obtained by the X-ray spectral analysis are consistent with those expected by gas sloshing. In addition, we find that the equation of state of disturbed gas is consistent with being isobaric, i.e., the motion of gas sloshing is in pressure equilibrium. The morphology of this dipolar patter seems to be a spiral. If so, the direction of the sloshing motion is in the plane of the sky. Such dipolar spiral patterns are often found in local cool core clusters \citep[e.g.,][]{Churazov03, Clarke04, Sanders14, Ueda17}, which show no apparent feature of a major merger. Those indicate that their gas sloshing is induced by a minor merger. \rxj, however, has a subcluster and the excess hot ICM in its central 150\,$h^{-1}$\,kpc. \rxj ~is therefore the first cluster ever known to host both a major merger and gas sloshing in the cool core.

The size of the dipolar pattern is a factor of $2 \sim 3$ smaller than that often found in local cool core clusters \citep[e.g.,][]{Ueda17}. Since the size tends to evolve with time after the passage of a subcluster \citep[e.g.,][]{ZuHone10}, gas sloshing in this cluster is expected to have occurred recently. The dipolar spiral pattern indicates that the orbit of the major merger is in the plane of the sky.  This is in accord with the fact that the redshift difference of the 2nd BCG from the BCG is less than 100\,km\,s$^{-1}$ by optical observations \citep{Lu10}. The stripped gas locates behind the subcluster, which indicates the infalling subcluster is in the first passage. It is unclear, however, whether or not the infalling subcluster is the origin of gas sloshing. If the direction of the passage is from southwest to northeast, it seems to be hard to disturb and create a rotational motion of the gravitational potential well of the main cluster. A possibility that the subcluster is in the first passage and gas sloshing is induced by another earlier merger, was suggested by \cite{Kreisch16}. In this paper, we can not constrain their scenario.

\subsection{Self-interaction cross section for dark matter}

As discussed above, it seems that the merger in \rxj ~is in the plane of the sky. Following \cite{Markevitch04}, an offset in the positions of galaxies, DM halos, and the ICM allows us to constrain the self-interaction cross section of DM. We estimated the surface mass density of DM ($\Sigma_{s}$) in the subcluster using its mass distribution shown in the left panel of Figure~\ref{fig:SL}. We measured the mean $\Sigma_{s}$ within a radius of 25\,$h^{-1}$\,kpc from the second mass peak. We also subtracted the contribution of the main cluster based on the parameters listed in Table~\ref{tab:SL}. Assuming the scattering depth of $\tau_{s} = \Sigma_{s} \times \sigma_{\rm DM} / m < 1$ for DM, we obtain the upper limit of the self-interaction cross section for DM to be $\sigma_{\rm DM} / m < 2.1\,h^{-1}$\,cm$^2$\,g$^{-1}$, where $\sigma_{\rm DM}$ is the DM collision cross section and $m$ is the mass of a DM particle. To derive more conservative upper limit, we further considered the uncertainty of the surface mass density map of \rxj. We subtracted the $2\sigma$ value of the rms noise obtained in Section~\ref{sec:hst} from the original surface mass density map and re-estimated the cross section in the same manner as mentioned above. The $2\sigma$ (95\% CL) upper limit of the cross section is then $\sigma_{\rm DM} / m < 3.7\,h^{-1}$\,cm$^2$\,g$^{-1}$. The derived upper limits are comparable to those reported in the studies of other clusters \citep[e.g.,][]{Markevitch04, Harvey15}.

\section{Conclusions}

We have studied \rxj, one of the well-known major merging clusters, combining the high angular resolution, multi-wavelength data taken by \Chandra, ALMA, and \HST. The conclusions of this paper are summarized as follows.

\begin{itemize}
\item The residual image of the X-ray surface brightness shows a clear dipolar pattern in the cluster center, whereas, we find no excess SZE signal in the central region in the SZE residual image. The dipolar pattern indicates that a fraction of the gas in the cool core is disturbed by gas sloshing. We have estimated the equation of state of the perturbation in gas using the X-ray and SZE residual images. We find that the inferred velocity is $420^{+310}_{-420}$\,km\,s$^{-1}$, which is much lower than the adiabatic sound speed of the 10\,keV ICM inside the core; thus, the perturbation is consistent with being isobaric. This is the first direct evidence of sub-sonic nature of gas sloshing motion. 

\item Both X-ray and SZE residual images show an excess component in the southeast substructure. We find that the peak of excess hot ($\sim 30$\,keV) ICM is located at $27''$ ($109\,h^{-1}$\,kpc) to the southeast from the cluster center. This region is faint in X-rays but bright in the SZE. The X-ray inferred thermal pressure of the excess component is nearly constant among the regions where the SZE signal is prominent. The velocity inferred from the equation of state of the excess hot ICM is $1970 \pm 150$\,km\,s$^{-1}$, which is comparable to the adiabatic sound speed of the 20\,keV ICM. This result supports a picture that the perturbation in the southeast is generated by shock.

\item The mass distribution of \rxj ~obtained with SL is reproduced well by the two dark halo components. Their mass peaks are in good agreement with the positions of the BCG and the 2nd BCG. The mass peak of the main cluster matches the X-ray centroid, while the mass peak of infalling subcluster is offset from the substructure in X-rays and the SZE. This indicates that the gas originated in the subcluster is stripped by ram-pressure of the main cluster and shock heated during an on-going major merger. 

\item In our scenario, this major merger is likely in the first passage. \rxj ~is therefore an exceptional cluster in which the excess hot gas, on-going major merger, and the sloshing cool core coexist within the central $150$\,$h^{-1}$\,kpc. 

\item We have also constrained the self-interaction cross section of DM. The resulting upper limit of the cross section is $\sigma_{\rm DM}/m < 3.7\,h^{-1}$\,cm$^{2}$\,g$^{-1}$ (95\% CL).

\end{itemize}

\acknowledgments
We are grateful to the anonymous referee for helpful suggestions and comments. This paper makes use of the following ALMA data: ADS/JAO.ALMA\#2013.1.00246.S. The scientific results of this paper are based in part on data obtained from the Chandra Data Archive: ObsID 506, 507, 3592, 13516, 13999, and 14407.  ALMA is a partnership of ESO (representing its member states), NSF (USA) and NINS (Japan), together with NRC (Canada) and NSC and ASIAA (Taiwan), in cooperation with the Republic of Chile. The Joint ALMA Observatory is operated by ESO, AUI/NRAO and NAOJ. The National Radio Astronomy Observatory is a facility of the National Science Foundation operated under cooperative agreement by Associated Universities, Inc. This work was supported by the Grants-in-Aid for Scientific Research by the Japan Society for the Promotion of Science with KAKENHI Grant Numbers JP24340035 (Y.S.), JP25400236 (T.K.), JP26400218 (M.T.), JP26800093 (M.O.), JP15H02073 (R.K.), JP15H03639 (T.A.), JP15H05892 (M.O.), JP15K17610 (S.U.), JP15K17614 (T.A.), JP16K05295 (N.O.), JP16H07086 (S.T.), and JP17H06130 (K.K. \& R.K.). This work was supported by NAOJ ALMA Scientific Research Grant Numbers 2017-04A. In addition, this work is supported in part by the Ministry of Science and Technology of Taiwan (grant MOST 106-2628-M-001-003-MY3).

\vspace{5mm}
\facilities{CXO, ALMA, HST}


\bibliographystyle{apj}
\bibliography{00_BibTeX_library}

\end{document}